\documentclass[aps,pra,preprint]{revtex4-2}
\usepackage{amsmath, amssymb}
\usepackage{xfrac}
\usepackage{placeins}
\usepackage{booktabs}
\usepackage{subcaption}
\usepackage[symbol*]{footmisc}
\usepackage{setspace}
\usepackage[normalem]{ulem}
\usepackage{xcolor}
\usepackage{physics}
\usepackage{lineno}
\usepackage{tikz}
\usetikzlibrary{quantikz}

\definecolor{darkgreen}{rgb}{0.05, 0.5, 0.06}

\newcommand{\btheta}{\boldsymbol{\theta}}

\begin{document}

\title{Considerations for evaluating thermodynamic properties with hybrid quantum-classical computing work-flows}

\author{Spencer T. Stober}
\email{spencer.t.stober@exxonmobil.com}
\author{Stuart M. Harwood}
\author{Dimitar Trenev}
\affiliation{ExxonMobil Corporate Strategic Research, Annandale, NJ 08801, USA}

\author{Panagiotis Kl. Barkoutsos}
\affiliation{IBM Quantum, IBM Research Europe, S{\"a}umerstrasse 4, 8803 R{\"u}schlikon, Switzerland}

\author{Tanvi P. Gujarati}
\affiliation{IBM Quantum, IBM Research Almaden, San Jose, CA 95120, USA}

\author{Sarah Mostame}
\affiliation{IBM Quantum, IBM T.J. Watson Research Center, Yorktown Heights, NY 10598, USA}

\date{\today}

\begin{abstract}
Quantum chemistry applications on quantum computers currently rely heavily on the variational quantum eigensolver (VQE) algorithm. This hybrid quantum-classical algorithm aims at finding ground state solutions of molecular systems based on the variational principle.
VQE calculations can be systematically implemented for perturbations to each molecular degree of freedom, generating a Born-Oppenheimer potential energy 
surface (PES) for the molecule.
The PES can then be used to derive thermodynamic properties,
which are often desirable for applications in chemical engineering
and materials design.
It is clear from this process that quantum chemistry
applications contain a substantial classical
computing component in addition to steps that can 
be performed using a quantum computer.
In order to design efficient work-flows that take full advantage of each hardware-type, it is critical to consider the entire process
so that the high-accuracy electronic energies possible
from quantum computing are not squandered in the process of calculating 
thermodynamic properties.
We present a summary of the hybrid quantum-classical work-flow
to compute thermodynamic properties.
This work-flow contains many options that can significantly 
affect the efficiency and the accuracy of the results,
including classical optimizer attributes, 
number of ansatz repetitions, and how the vibrational Schr{\"o}dinger equation is
solved to determine vibrational modes.
We also analyze the effects of these options by employing robust statistics along with simulations 
and experiments on actual quantum hardware.
We show that through careful selection of work-flow options, 
nearly order-of-magnitude increases in accuracy are possible
at equivalent computing time.
\end{abstract}

\maketitle

\section{Introduction}

Thermodynamic properties of molecules and their mixtures are vital to 
process modeling, design, and control \cite{kontogeorgis_2021}.
These properties can sometimes be determined experimentally, however
this is often difficult or expensive, which makes quantum mechanical
prediction of thermodynamic properties attractive.
Gas phase processes, such as thermal cracking, pyrolysis, and combustion,
are particularly amenable to modeling with these methods, 
and development of theories and software to carry this out is an active area of research \cite{green_2020}.
 
Calculation of thermodynamic properties from quantum mechanical calculations typically requires several steps: 
(i) calculation of the Born-Oppenheimer potential energy surface (PES);
(ii) determination of intramolecular vibrational modes; and
(iii) calculation of the partition function and thermodynamic observables as a function of temperature and pressure.
Typically, the first step is a computational bottleneck since it requires solving the electronic Schr{\"o}dinger equation for several sets of fixed nuclear coordinates that span relevant conformations of the molecular system on its PES. 

Quantum computers provide a new framework that can be useful for this first step and have potential to accelerate this task compared to classical computers \cite{abrams_1997, kassal_2011, omalley_2016, kandala_2017, barkoutsos_2018, hempel_2018, romero2018, babbush2018, cao_2019, ollitrault_2020,gao2021computational,jrice_2021}. 
In addition, within a selected basis set the solution that could be obtained by a quantum calculation is the full configuration interaction (FCI) result because all possible combinations of the single particle basis functions \cite{abrams_1999} can be included in the most general case.
Indeed, developing practical methods to approach FCI-accuracy is a goal of substantial research in theoretical chemistry
(see also review of Ref. \citenum{bartlett_2007}). 
If quantum computers are able to achieve this for systems of relevant size, it will represent a paradigm shift in computational chemistry.

However, FCI-accuracy electronic structure calculations must also be accompanied by improved downstream work-flows that take advantage of these calculations when thermodynamic property predictions are required. 
Such is the case, for example, in modeling reaction kinetics and process engineering. 
Areas that will need to be considered towards such an effort are approaches to solving the vibrational Schr{\"o}dinger equation (to obtain the molecular vibrational energy levels), and statistical mechanics methods (to compute thermodynamic properties).

Using classical computers to compute PESs has evolved over decades, but using quantum computers to aid in this task is a new area of research\cite{omalley_2016, kandala_2017, hempel_2018, colless_2018}. 
Traditionally, computational chemists rarely needed to worry about reproducibility, since on classical computers the results of a calculation are deterministic.
However, the statistical noise inherent to quantum computing algorithms means that each measurement of an electronic energy
will not necessarily be identical.
This is even more the case on the noisy quantum devices available today, where certain experiment parameters can  lead to quantum circuits that are wide (many
qubits) or deep (circuits  consisting of many operations that can take a long time to execute relative to the decoherence time of qubits) and can give rise to errors with broad distributions\cite{preskill2018quantum, roffe:19}.
These noisy predictions of electronic energy are then propagated into the PES, the vibrational energy levels, and ultimately, the predicted thermodynamic properties. 

For computing electronic energies, the variational quantum eigensolver (VQE) is a promising quantum algorithm that can be implemented on near-term noisy quantum computers since it can approximate solutions to  the Schr{\"o}dinger equation using a hybrid quantum-classical approach that requires only few qubits and
relatively shallow circuits \cite{peruzzo_2014,kandala_2017}.
Other approaches, such as the quantum phase estimation algorithm  \cite{aspuru_guzik_2005} can also be used to approximate solutions to the Schr{\"o}dinger equation.
Unfortunately, the high requirements in terms of gate operations (and consequently decoherence time) to implement this algorithm limits its application on currently available hardware \cite{reiher2017elucidating}. 
For this reason, we will focus on the VQE algorithm in the forthcoming discussion.

When implementing the VQE algorithm, the molecular system is described with a fermionic Hamiltonian in the second-quantized form using one- and two-body integrals in a specified basis set. This fermionic Hamiltonian is then properly mapped into a qubit Hamiltonian using appropriate fermion-to-qubit transformations\cite{jordan-wigner_1928, bravyi_2002, bravyi_2017} . Estimation of the expectation value of such a qubit Hamiltonian for a trial quantum state provides the energy of the molecular system with respect to the quantum state. In the VQE algorithm, a trial quantum state ansatz is created as a quantum circuit with quantum gates (typically decomposed into single and two qubit operations).
An ansatz contains tunable circuit parameters which are optimized classically during the VQE algorithm such that the energy of the system evaluated on a quantum processor with respect to the ansatz is minimized. Increasing the number of these tunable parameters can increase the flexibility of the ansatz to represent the ground-state wavefunction \cite{guzik_2019} describing the system under consideration.

There are two main approaches for constructing an ansatz to represent the target wavefunction.
The ansatz can be constructed based on physical insight of the electronic structure, such as the Unitary Coupled Cluster (UCC) ansatz \cite{peruzzo_2014, romero2018, barkoutsos_2018} or the variations of the method \cite{Sokolov2020, grimsley2018adapt, tang2021qubit}, that exploit coupled-cluster theory to describe the wavefunction \cite{bartlett_2007}.
Unfortunately, the implementation of UCC ansatz on quantum  hardware requires quantum circuits with excessive circuit depth
that cannot be executed effectively on near-term quantum hardware 
(although, UCCSD ansatz based VQE experiments have been demonstrated for HeH+ ion using trapped-ion quantum computing architecture \cite{shen_2017}).
To account for this limitation, Kandala et al. \cite{kandala_2017} proposed the second popular approach, an ansatz that can be constructed by leveraging the native basis gate set and topology of the selected quantum computer.
In this way the depth of the circuit can be significantly reduced to the point that can  be executed on currently available hardware.

Being able to calculate the ground state energies for different molecular geometries gives us access to the PES and eventually to the computation of vibrational energy levels. 
Computing vibrational energy levels from the PES can be approached in several ways.
The least complex way is by choosing a potential with analytical eigenvalues (e.g., the harmonic or Morse \cite{morse_1929} potential)  and fitting it to each degree of freedom on the PES.
The eigenvalues will then correspond to the intramolecular vibrational modes.
However, this method suffers from two main drawbacks: the PES may not be accurately represented by the  chosen potential, and it assumes that the degrees of freedom are uncoupled.
In fact, if we eliminate these drawbacks by fitting an arbitrary functional form to the PES (e.g., a  spline) and use a numerical eigensolver to compute the vibrational modes of this fully coupled surface, it is equivalent to solving the vibrational Schr{\"o}dinger equation for the vibrational degrees of freedom.
This, however, leads to the same issues with scaling as are prevalent in the electronic Schr{\"o}dinger equation  (although, vibrational calculations remain tractable for slightly larger molecules than the electronic ones since the number of electrons increases much faster than the number of nuclei). 
The calculation of the vibrational energy levels constitutes an important step
since they are the main input to the vibrational partition function,
which allows computing thermodynamic properties of molecules
(more details on this are discussed later).

In this work we re-evaluate the overall hybrid quantum-classical computing work-flow for calculation of thermodynamic properties, and highlight some of the more common approximations and potential accuracy losses.
This includes calculating electronic energies, construction of the PES, solving the vibrational Schr{\"o}dinger equation, and using statistical mechanics to compute thermodynamic properties.
We use the VQE algorithm and devise a work-flow that allow the results to be reproducible within the statistical uncertainty of the quantum computation. This work-flow is demonstrated for the LiH molecule modelled using the minimal basis. Finally, we discuss applying our findings to computing the PES of LiH on the ibmq$\_$rome \cite{Qiskit} quantum computer with IBM Quantum Falcon processor
\textemdash a superconducting transmon quantum device with 5-qubits. 
We show that by careful selection of work-flow options, it is possible to obtain high-accuracy thermodynamic properties based on VQE-calculated electronic energies and a numerical solution of the vibrational  Schr{\"o}dinger equation.

\section{Theory and Methods}

\subsection{Variational quantum eigensolver}
The variational quantum eigensolver algorithm introduced in Ref.~\citenum{peruzzo_2014} is a hybrid quantum-classical algorithm based on the variational principle, which is used to approximate the lowest eigenvalue
of a given Hamiltonian, $\mathcal{H}$. 
The procedure comprises choosing a parameterized 
wavefunction, $ |\psi\left(\boldsymbol{\theta}\right)\rangle$, to minimize the expectation value $E_\mathcal{H}(\psi(\boldsymbol{\theta}))=\expval{\mathcal{H}}{\psi(\boldsymbol{\theta})}$. 
This wavefunction is represented by a parameterized quantum circuit and a quantum computer is used to efficiently estimate the expectation value.
Note that, since quantum states are normalized, $E_\mathcal{H}(\psi)$ is just the Rayleigh quotient of the complex Hermitian matrix, $\mathcal{H}$, and a non-zero vector $\psi$. 
It is clear that the lowest possible value for $E_\mathcal{H}$ is the smallest eigenvalue, $\lambda_0$, of $\mathcal{H}$ obtained when $|\psi\rangle = |\psi_0\rangle$ is the corresponding eigenstate. 
The set of parameters $\boldsymbol{\theta}$, also called variational parameters, are varied through a classical optimization loop until the expectation value converges to the minimum. 

The simplicity of the theory behind the VQE algorithm is somewhat deceptive. 
In practice, to guarantee that one has indeed succeeded in accurately computing $\lambda_0$, two conditions must hold. 
First, whatever ansatz one uses needs to be good enough to obtain (or, at the very least, reasonably approximate) the eigenstate $|\psi_0\rangle$, which is a-priori unknown. 
Second, assuming that there are indeed parameters $\btheta_0$ such that $|\psi(\btheta_0)\rangle\approx|\psi_0\rangle$, the minimization algorithm used needs to be robust enough to find those optimal parameters. 
As a further complication, without using some application-specific knowledge, it might be hard to satisfy both of these conditions at once, as they are, in a sense, at odds with each other. 
For example, increasing the dimension of the parameterization space, and thus making the ansatz more expressive, might make it easier to satisfy the first condition, but it almost certainly makes satisfying the second condition harder. 
The importance of choosing the right ansatz and optimizer is paramount for the success of VQE in practice.

\paragraph{Hamiltonian mapping and trial wavefunctions for LiH.}
The first step in computing the electronic energy 
is the specification of the basis set and fermion-to-qubit mapping scheme to represent the electronic structure Hamiltonian. 

For the case of LiH, there are six spatial orbitals in STO-3G basis set, out of which five belong to Li (1s, 2s, 2px, 2py and 2pz), and one to H (1s). 
These 6 spatial orbitals correspond to a total of 12 spin orbitals. We assume that the core orbital, 1s, of Li is kept frozen at the Hartree-Fock level which is a standard practice when using minimal basis sets. 
To further reduce the number of qubits required to represent the molecule, we freeze orbitals 2py and 2pz on Li (for LiH oriented along x axis) at the Hartree-Fock level.
An average error of ~0.7mHa across the LiH bond dissociation profile was obtained on freezing these two orbitals for the range of bond lengths considered in this paper
(compared to results without freezing the orbitals).
Using parity\cite{bravyi_2017} fermion-to-qubit mapping on the remaining 6 spin orbitals, each mapped qubit now represents the parity, i.e., presence of odd or even number of particles up to the spin orbital in consideration.
Because the operators corresponding to total number of spin up electrons and total number of spin down electrons commute with the Hamiltonian, the number of spin up and spin down electrons are independently conserved. 
In the parity mapping, the qubits that represent the sum of all spin up electrons and the sum of all electrons (spin up + spin down) remain unchanged during the calculation and can be removed. 
Therefore, we can reduce the number of qubits required from 6 to 4 by 
using the intrinsic properties of the parity fermion-to-qubit mapping \cite{bravyi_2017}.

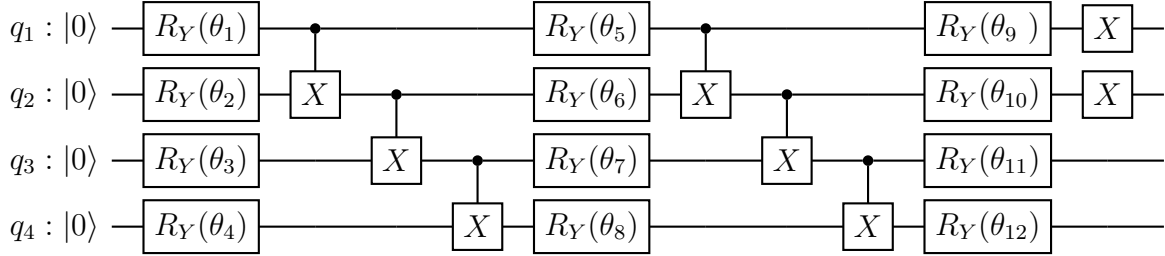
\begin{figure}[h]
\centering
\begin{tikzpicture}
\node[anchor=north] (A1) at  (1.8,0) {
\begin{quantikz}[column sep=12pt, row sep={25pt,between origins}]
\lstick{$q_1: \ket{0}$} &\gate{R_Y(\theta_1)}  &\ctrl{1} &\qw      &\qw      &\gate{R_Y(\theta_5)} &\ctrl{1} &\qw      &\qw      &\gate{R_Y(\theta_{9 \ })}  &\gate{X} &\qw\\
\lstick{$q_2: \ket{0}$} &\gate{R_Y(\theta_2)}  &\gate{X} &\ctrl{1} &\qw      &\gate{R_Y(\theta_6)} &\gate{X} &\ctrl{1} &\qw      &\gate{R_Y(\theta_{10})} &\gate{X} &\qw\\
\lstick{$q_3: \ket{0}$} &\gate{R_Y(\theta_3)}  &\qw      &\gate{X} &\ctrl{1} &\gate{R_Y(\theta_7)} &\qw      &\gate{X} &\ctrl{1} &\gate{R_Y(\theta_{11})} &\qw      &\qw\\
\lstick{$q_4: \ket{0}$} &\gate{R_Y(\theta_4)}  &\qw      &\qw      &\gate{X} &\gate{R_Y(\theta_8)} &\qw      &\qw      &\gate{X} &\gate{R_Y(\theta_{12})} &\qw      &\qw
\end{quantikz} };
\end{tikzpicture}
\caption{\label{fig_lih_circuit} Schematic diagram of the parameterized circuit (ansatz) using two repetitions of $R_Y$ gates 
and four qubits which are linearly entangled with controlled-$X$ gates. The application of the last two X-gates allow us to construct the Hartree-Fock (HF) state in a meaningful way for a specific initialization of all the circuit parameters.}
\end{figure}

Following the approach introduced in Ref.~\citenum{kandala_2017}, the trial wavefunction can be represented by 
\begin{equation}
\ket{\psi(\btheta)} = 
U_{HF}  U_{N+1}(\btheta) 
U_{ENT} U_{N}(\btheta)
\dots 
U_{ENT} U_1(\btheta)
\ket{0\dots0}
\label{eq:ansatz}
\end{equation}
where
the $U_i(\btheta)$ are parameterized unitary operations and $U_{ENT}$ is a block of entangling gates consisting of two-qubit operations.
Layers of $U_i(\btheta)$ and $U_{ENT}$ are repeated $N$ times and a last block of parameterized operations $U_{N+1}(\btheta)$ is applied.
This is followed by $U_{HF}$, a unitary that 
maps the $\ket{0...0}$ state to the initial Hartree-Fock state.
Figure~\ref{fig_lih_circuit} shows the form of these operations in more detail;
each $U_i(\btheta)$ operation consists of parameterized $R_Y$ gates and only depends on a subset of the variational parameters, while the entanglement block $U_{ENT}$ consists of controlled-$X$ (cNOT) operations between pairs of qubits.
$U_{HF}$ is implemented with the two final $X$ gates on the first and second qubits.

Note that when all the variational parameters are set to zero, this variational ansatz produces the Hartree-Fock state, $U_{HF} \ket{0\dots 0}$.
While it is possible to place $U_{HF}$ at the beginning of the circuit (as the first operator), the action of the
controlled-$X$ gates would alter the state even when variational parameters are set to zero.
It is less obvious what the resulting state is, or how to set the ansatz parameters so that a meaningful reference state is obtained when using a heuristic trial wavefunction.
Knowing precisely how to prepare the Hartree-Fock state proves valuable to improving the accuracy of the ground state estimates and eventually for the calculation of the PES.

\paragraph{Classical optimizer}
The choice of the classical optimizer is of paramount importance in the VQE algorithm, as shown in Refs.~\citenum{lavrijsen2020classical, harwood_2021}.
In this work we use stochastic gradient descent (SGD), 
following the implementation and augmentations to the method discussed in Ref.~\citenum{harwood_2021}. 
The details related to the implementation of the method and a summary of the techniques is provided in Appendix~\ref{app:VQE_classical_opt}.

As a local optimization method, the performance of SGD is very sensitive to the initial value of the ansatz parameters $\boldsymbol{\theta}$.
However, in the broader context of applying VQE to calculate a PES, there are techniques to avoid this limitation.
See Section~\ref{sec:computes_pes} for a detailed discussion.

\FloatBarrier
\subsection{Accurately computing the PES}
\label{sec:computes_pes}
Although the PES is simply a collection of ground state electronic energies for different molecular geometries, calculating it with VQE and the noisy energy estimates that come from a quantum computer pose certain challenges.
To account for the effect of noise and statistical uncertainty, and to boost the performance of the VQE algorithm we apply certain techniques described below. 

\paragraph{Bootstrapping}
As a gradient-based method, the performance of SGD can be very sensitive to the choice of initial ansatz parameters.
A poor choice of initial parameters can cause SGD to converge to a local minimum of the objective function $E_H(\psi(\cdot))$ that does not approximate the smallest eigenvalue.
In the context of computing a PES, however, the sequence of optimization problems that result from applying VQE at various molecular degrees of freedom are related.
One can take advantage of this to use previously determined optimal ansatz parameters as the initial choice in a new application of VQE.
In this work, we simply look at the closest previously computed molecular degree of freedom (if one exists) and use the corresponding optimal ansatz parameter values as the initial guess for the current one.

\paragraph{Initialization}
To begin this bootstrapping procedure, we start the calculation of the PES at the equilibrium geometry and set the initial 
ansatz parameters to zero.
As discussed earlier, when the ansatz parameters are set to zero our choice of ansatz produces the Hartree-Fock state.
At the equilibrium geometry, the Hartree-Fock state is a reasonable approximation to the ground state, since electron-electron
correlations are lower.
In this way, we begin the calculation of the PES with good ansatz parameters, and subsequently we move outward towards shorter and longer bond lengths so that the bootstrapping procedure continues to use high-quality initial ansatz parameters.

\paragraph{Resampling}
One final consideration is how to obtain an accurate estimate of the final, optimized expectation, which is our estimate of the minimum eigenvalue.
While SGD can converge with very noisy gradient estimates, this level of noise is often unacceptable for the final PES.
Consequently, one can follow the procedures in Ref.~\citenum{kandala_2017} (Section V.A) or Ref.~\citenum{mcclean2016theory} to estimate the variance of the estimator of $f = E_\mathcal{H}(\psi(\cdot))$.
Assuming this variance is $\sigma^2$, we can get an estimate of $f$ with standard deviation less than or equal to $\varepsilon_r$ by taking the average of $N_r$ repeated independent estimators, with
$N_r > \sfrac{\sigma^2}{\varepsilon_r^2}$.
This resampling only needs to be applied \emph{once} for each molecular geometry, after the optimized ansatz parameters have been found.

\subsection{Calculating Thermodynamic Observables}
\label{sec:thermo_observables}
For the calculation of thermodynamic observables we need to estimate the vibrational energy levels and the system partition function. 
Once the partition function is acquired the calculation of the internal energy of the system and the enthalpy can be estimated. 
In this section we elaborate on the classical work-flow from the calculation of the PES to the thermodynamic observable calculation.

\paragraph{Computing vibrational energy levels of the PES.}

The vibrational Schr{\"o}dinger equation is given as:
\begin{equation}
\left( -\frac{\hbar^2}{2m_{R}} \: \laplacian + V(r) \right) \:
\Psi_n(r) = \epsilon_n \: \Psi_n(r),
\end{equation}
where $m_{R}$ is the relevant reduced mass, $\hslash$ is the reduced Planck's constant, $V(r)$ is the potential energy function representing the PES, and $\Psi_n(r)$ represents the $n^{th}$ eigen-function as a function of the positions of the nuclei.
Our goal is to find the eigenvalues, $\epsilon_n$, of this equation for a given functional form of PES. The obtained eigenvalues are then used in the computation of the vibrational partition function and eventually, thermodynamic properties.

Fitting the PES with different functional forms can lead to significant differences in the calculation of the vibrational energy levels. 
Using the harmonic potential, the Morse potential, and an arbitrary cubic spline fit to get predicted vibrational modes, and subsequently predicted thermodynamic properties can illustrate these variations. 

A detailed description of molecular vibrational modes obtained from harmonic, Morse and cubic spline fits of PES can be found in Appendix~\ref{app:vibrational} and Ref.~\citenum{book_wilson}.

\paragraph{The partition function.}
The partition function for a molecule is the product of contributions from translational, rotational, vibrational, electronic, and nuclear degrees of freedom, as shown in equation \ref{eq_mol_pf}, where $V$ and $T$ are volume and temperature, respectively.
\begin{equation}
q(V,T) = 
q_{translation} \:
q_{rotation} \:
q_{vibration} \:
q_{electronic} \:
q_{nuclear}.
\label{eq_mol_pf}
\end{equation}
We assume that the electronic and 
nuclear contributions are unity since, for LiH, we do not
consider excited electronic or nuclear states.

The calculation of the individual contributions
to the molecular partition function are described 
in detail in Ref.~\citenum{book_mcquarrie} and details pertinent to our analysis are highlighted in Appendix~\ref{app:partition_function}.

\paragraph{Thermodynamics.}
Once $q(V,T)$ has been determined as a function of temperature, the internal energy, $U$, can be obtained via differentiation (at constant $V$)
\begin{equation}
U
= k_B \: T^2  
\left. 
\frac{\partial \ln q(V,T)}{\partial T}
\right|_{V}.
\end{equation}
The enthalpy, $H$, and constant-pressure heat capacity, $C_p$, are then easily computed as
\begin{equation}
H = U + PV
\quad\text{and}\quad
C_p = \left. \frac{\partial H}{\partial T}\right|_{N,P},
\end{equation}
where $PV = k_B \, T$.

\FloatBarrier
\section{Results and Discussion}

In the following section we present the attributes of the calculations performed with the hybrid quantum-classical work-flow for estimating the thermodynamic properties of LiH. We analyze the requirement of resources when the problem is executed on a quantum computer, i.e., the number of ansatz repetitions in the quantum circuit and how many measurement shots are required to achieve reliable results. In addition, we also compare the three functional forms of the potential, discussed in Section \ref{sec:thermo_observables}, and how they affect the accuracy of the vibrational energy levels and calculated thermodynamic properties.

For the resource analysis we use simulations of quantum devices in lieu of actual hardware using the \textit{qasm\_simulator} available via Qiskit\cite{Qiskit}. The finite sampling of the probability distribution of possible outcomes that results from measuring the qubits after the circuit has been executed introduces statistical noise (that we refer to as \emph{shot noise}).
Within this particular framework, we investigate the effect of different approaches (bootstrapping, resampling) used with the VQE algorithm and consequently for the calculation of the PES. 
Experiments performed on real quantum computing devices also suffer from additional types of noise beyond the statistical noise. 
In consideration of near-term noisy devices, we also evaluate and validate our work-flow on IBM Quantum hardware. 
The results obtained from simulations and hardware experiments provide a good benchmark for the calculation of thermodynamic observables implemented using the hybrid quantum-classical work-flow.

\FloatBarrier
\subsection{The effect of ansatz repetitions and shots on the PES}
A key step in using a quantum computer 
to perform VQE is to determine the operating parameters under which results are reliable and the calculation is efficient.
Since measurements on quantum computers are probabilistic and subject to noise, analyzing the number of circuit evaluations (shots) used to evaluate expectation values is vital to achieve accurate results. 
At the same time, the ability of the ansatz to represent the molecular system is also important to obtain meaningful results. 
As discussed earlier, the entangling and $R_Y$ blocks of the ansatz chosen in this study can be replicated multiple times to increase the number of parameters used in VQE, but this makes it more difficult for the optimizer to converge to a global minimum. 

\begin{figure}[h]
\centering \includegraphics[width=0.9\textwidth]{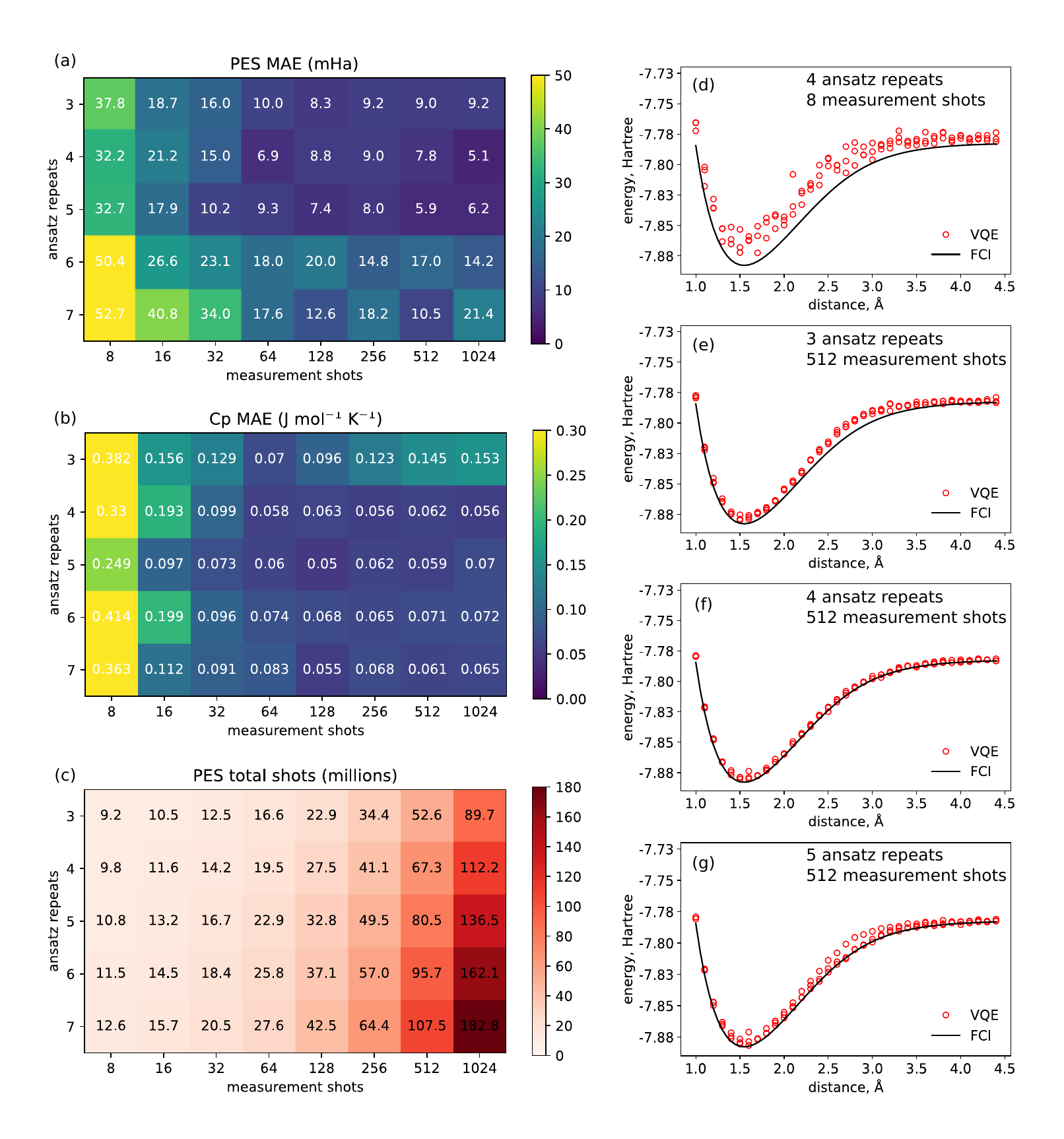}
\caption{\label{fig_mae_vs_shots_composite} 
(a) the MAE of the PES as a function of ansatz repetitions and measurement shots.
(b) shows the computed $C_p$ for each PES. 
(c) is the total number of shots used in each calculation, which includes shots used in the resampling procedure.
(d-g) representative PESs corresponding to selected numbers of measurement shots and
ansatz repeats.
}
\end{figure}

Figure~\ref{fig_mae_vs_shots_composite}(a) shows the mean absolute error (MAE) obtained over 5 independent batches of the PES for different combinations of ansatz repetitions and shots. Each batch consists of 3 repeats of VQE energies for every geometry, which were then averaged.
For the VQE procedure, the SGD classical optimizer algorithm
was used with bootstrapping and resampling,
which will be discussed in detail later. 
More details on how the analysis was performed can be found in Appendix~\ref{sec_procedure_mae}.

For our 4-qubit ansatz, $N =$ 3 to 7 repetition of layers (refer Equation \ref{eq:ansatz}) corresponds to 16, 20, 24, 28, and 32 tunable variational parameters, respectively.
Figure~\ref{fig_mae_vs_shots_composite}(a) shows that using more than five ansatz repetitions results in higher MAE, which is likely due to the difficulty of optimizing more parameters. This illustrates an important point that increasing the complexity of the ansatz does not guarantee good VQE performance. 

In Figure~\ref{fig_mae_vs_shots_composite}(c), we show the total number of shots used to construct three replicates of the PES, averaged over the five batches.
Here as well, it is clear that as the number of ansatz repetitions increases and so does the number of variational parameters, and additional shots are needed by the optimizer to converge.

Figure~\ref{fig_mae_vs_shots_composite}(b) shows the MAE of the constant pressure heat capacity, $C_p$, with respect to exact reference calculations, for different combinations of ansatz repetitions and shots.
It is evident from this analysis that using too few shots results in poor accuracy for any number of ansatz repetition of blocks studied. In this case, at least 64 shots are necessary to achieve low MAE for the PES and $C_p$.

Comparing Figure~\ref{fig_mae_vs_shots_composite}(a and b) with Figure~\ref{fig_mae_vs_shots_composite}(c) we observe that beyond 64 shots, the MAE of the PES and $C_p$ does not improve substantially, but the number of total shots required to generate the PES grows rapidly. 

To illustrate the features of the PES obtained from VQE energies, we show representative PESs corresponding to selected number of ansatz repetitions and measurement shots in Figure~\ref{fig_mae_vs_shots_composite}(d-g).
We show the PESs for only one of the 5 batches for clarity.

Figure~\ref{fig_mae_vs_shots_composite}(d) corresponds to four ansatz repetitions (N=4) and eight measurement shots. This PES exhibits much larger MAE when compared to Figure~\ref{fig_mae_vs_shots_composite}(f) with 512 measurement shots. In Figure~\ref{fig_mae_vs_shots_composite}(e) we observe that with three ansatz repetitions and 512 measurement shots, the MAE is comparable with that obtained for four ansatz repetitions and same number of shots.
However, Figure~\ref{fig_mae_vs_shots_composite}(b) shows that the $C_p$ computed from the PES obtained with three ansatz repetitions exhibits higher MAE. In general the MAE of $C_p$ using three repetitions is higher than when four or more are used (with a sufficient number of measurement shots for each respective case).

Inspection of the PES in Figure~\ref{fig_mae_vs_shots_composite}(e), which corresponds to three ansatz repetitions shows a bulge around 2.5 - 3 \AA{} where the VQE-computation does not converge to the FCI value.
This observation holds for nearly all of the computed PESs with this combination of shots and ansatz repetitions. This is likely a result of the insufficient number of parameters available with three ansatz repetitions to represent the wavefunction of electronic configurations in this area of the PES.
This observation was also reported in Ref.~\citenum{kandala_2017} for a similar ansatz, but with a less efficient optimizer, reinforcing our conclusion that this is due to an insufficient number of variational parameters. Figure~\ref{fig_mae_vs_shots_composite}(g) also has a slight bulge in 2.5 - 3 \AA{} area, however this bulge is only present in about half of the computed cases. In view of the results for the four ansatz repetitions, we expect that the number of variational parameters existing in five or more repetition blocks should also be capable of representing the wavefunction with good accuracy. Therefore, the bulge seen intermittently with five ansatz repetitions is likely due to the optimizer struggling with more parameters, or the surface of the parameter space exhibiting local minima, which impede reliable convergence. 

The VQE-computed PES in Figure~\ref{fig_mae_vs_shots_composite}(f) with four ansatz repetitions and 512 measurement conforms well to the reference FCI calculation and the MAE of $C_p$ is also low. We conclude that four ansatz repetitions are likely sufficient to represent the wavefunction. We will use this combination for further discussion.

From these results, it is evident that the optimization of the calculation parameters (e.g. ansatz repetition blocks, number of variational parameters, number of measurements) is of critical importance for chemical calculations on near-term quantum computers. 
To derive the calculation parameters, exploratory experiments are needed.

\FloatBarrier
\subsection{Computing the PES by VQE}
The classical optimizer is a critical component of VQE. 
We now discuss the results obtained on augmenting SGD with bootstrapping and resampling,
which is thought to have an advantage for computing electronic energies with VQE compared to traditional optimizers \cite{stober_2020, harwood_2021}.

\begin{figure}
\centering \includegraphics[width=0.9\textwidth]{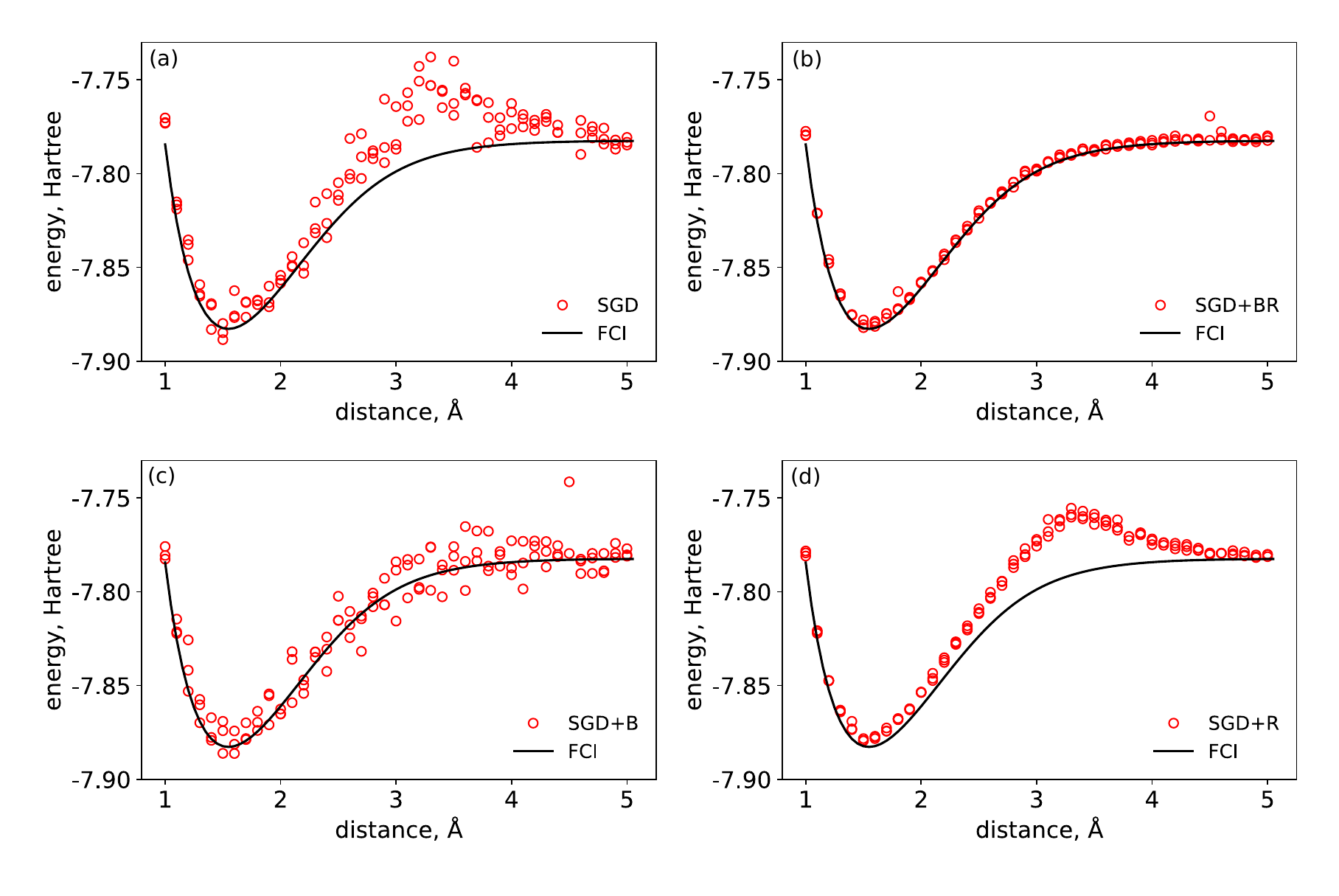}
    \caption{\label{fig_graber_bopes_compare}        
Comparison of four implementations of the VQE classical optimizer algorithm.  
(a) Steepest gradient descent without bootstrapping or resampling (SGD);
(b) with bootstrapping and resampling (SGD+BR);
(c) with bootstrapping only (SGD+B);
(d) with resampling only (SGD+R). 
The simulations include shot noise only and include four ansatz repetitions and 512 measurement shots.}
\end{figure}

The results shown in Figure~\ref{fig_graber_bopes_compare} support this conclusion. In this figure we report data from one of the five batches of 512 shots and 4 ansatz repetitions seen in Figure~\ref{fig_mae_vs_shots_composite}.
In Figure~\ref{fig_graber_bopes_compare}(a) the PES corresponds
to the SGD optimizer without bootstrapping or resampling,
and (b) corresponds to the baseline procedure we use with both resampling and bootstrapping included.
Figure~\ref{fig_graber_bopes_compare}(c) includes bootstrapping,
and since bootstrapping uses optimal parameters from the previous point, this information assists VQE in converging to the global minima (the ground state energy) rather than a local minima.

In the Figure~\ref{fig_graber_bopes_compare}(c) bootstrapping is enabled, but resampling is disabled. Here, the VQE algorithm is able to converge to the correct ground state energy but since the objective function is not sampled enough times to reduce its standard deviation, the reproducibility of each of the three replicates within that batch at each interatomic distance is poor.

Of particular importance is that, in Figure~\ref{fig_graber_bopes_compare}(a and d) the PES without bootstrapping exhibits a bulge around 2.5 to 4.0 \AA{}, which was also observed in Figure~\ref{fig_mae_vs_shots_composite}(g).
This is a clear indication that the limitation is not imposed by the number of repetition blocks, but is an artifact from the failure of the classical optimizer to converge without bootstrapping. 

In addition to increased accuracy, resampling and bootstrapping
improve the efficiency of the VQE algorithm. 
Table \ref{tab:pes_shots} shows the number of shots required
to construct the PESs discussed above and shown in 
Figure~\ref{fig_graber_bopes_compare}.
First, we see that because resampling is only performed
on the converged parameters of the ansatz, it adds a 
minimal number of shots to the total required to generate
the PES 
(i.e., the total number of shots required at each
iteration of VQE is large compared to the number of additional
shots required to resample the converged parameters).

Next, we see that in the cases of bootstrapping only (SGD+B) and 
with bootstrapping and resampling (SGD+BR) the
bootstrapping procedure reduces the number of shots
required to construct the PES.
This is a result of providing the optimizer a good initial
guess, which reduces the number of iterations required
to converge to the optimal ansatz parameters.

This study shows that including resampling and bootstrapping strategies together is important for getting an accurate and precise PES,
as shown by the lower MAE for SGD+BR in table \ref{tab:pes_shots}.
Furthermore, these procedures provide these improvements
while also decreasing the number of shots required to 
construct the PES.

\begin{table}[htbp]
  \centering
  \caption{
  Shots required to generate the example PESs shown
  in Figure~\ref{fig_graber_bopes_compare} using the listed 
  classical optimizer algorithms.
  Mean absolute errors (MAE) of the PESs are also shown.
  Confidence intervals are at the $ 95\% $ level and are determined
  from five replicates. 
  }
    \begin{tabular}{lcc}
    \toprule
    Optimizer & PES Shots (millions) & PES MAE (mHa) \\
    \midrule
    SGD   & \multicolumn{1}{c}{$77.8 \pm 2.3$} & $9.6 \pm 0.9$\\
    SGD+R & \multicolumn{1}{c}{$79.8 \pm 1.6$} & $7.9 \pm 0.2$\\
    SGD+B & \multicolumn{1}{c}{$62.6 \pm 1.7$} & $2.8 \pm 0.8$\\
    SGD+BR & \multicolumn{1}{c}{$66.4 \pm 0.6$} & $1.6 \pm 0.3$\\
    \bottomrule
    \end{tabular}%
  \label{tab:pes_shots}%
\end{table}%

\FloatBarrier
\subsection{Computing eigenvalues of the PES}
\label{sec_computing_eigvls}
\begin{figure}
    \centering    
     \includegraphics[width=0.6\textwidth]{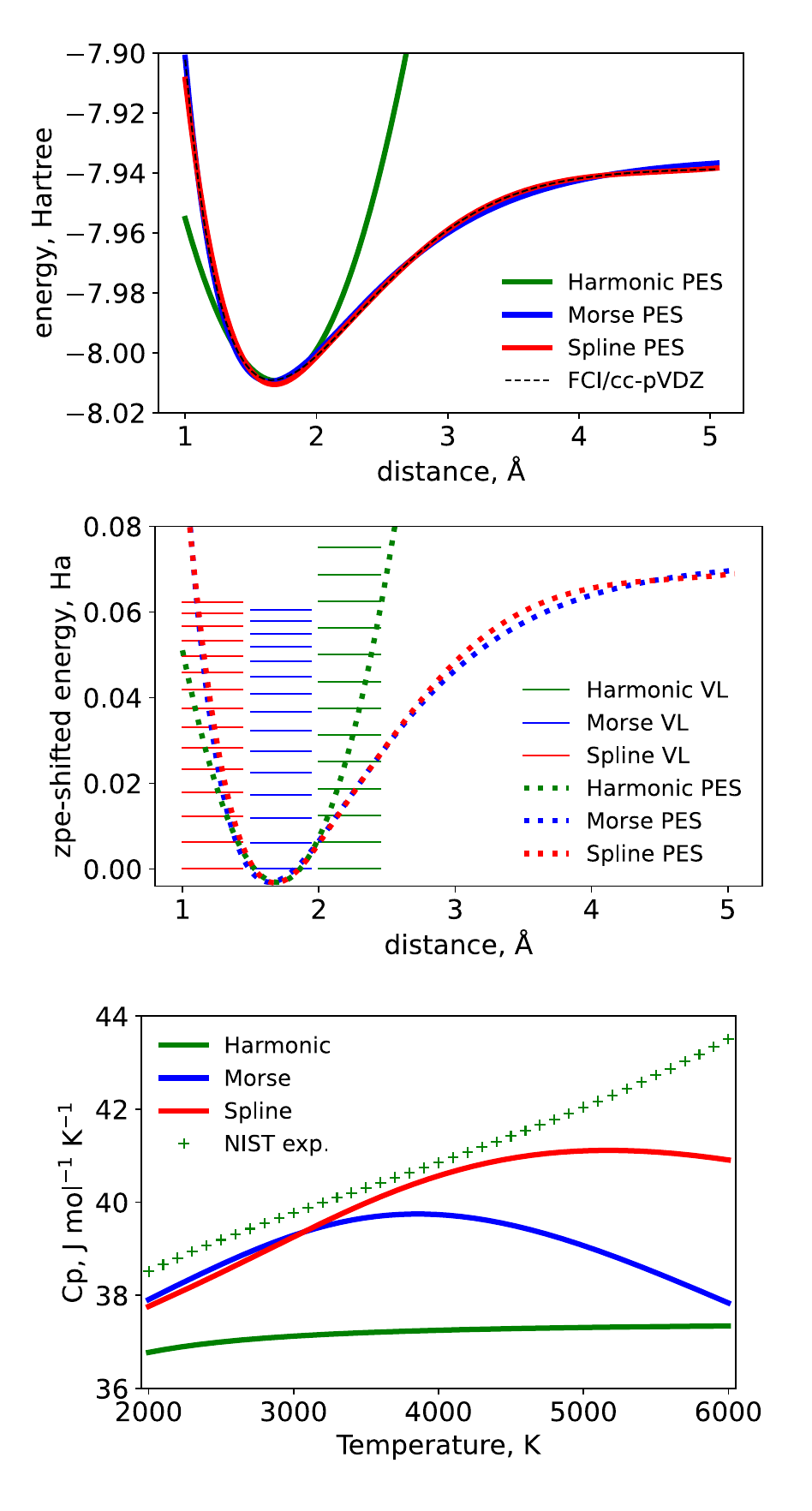}

    \caption{\label{fig_lih_pot_compare} 
    Comparison of harmonic, Morse, and spline fits of a classically computed FCI/cc-pVDZ PES.
    The top panel shows the fits of the PES.
    The middle panel shows vibrational energy levels as horizontal lines, which have been shifted to zero by their zero-point-energies.
    The bottom panel shows predicted heat capacity and experimental data as green plus-signs. }
\end{figure}

After computing the PES using the \textit{qasm\_simulator} we fit them to a functional form in order to be able to compute vibrational energy levels and thermodynamic properties. To illustrate the differences in methods for fitting the PES to a functional form and determining its eigenvalues, high-accuracy quantum mechanical calculations were performed on a classical computer. These were then used in our work-flow to compute vibrational energy levels and thermodynamic properties as an accurate benchmark.
For the LiH system, the PES was constructed at the FCI/cc-pVDZ level of theory \cite{dunning_1989} at intervals of 0.05 \AA{} from 1 to 5 \AA{} using Psi4\cite{psi4}.
The cc-pVDZ basis set was chosen since it can capture
the vibrational spectrum and geometry of the LiH system well 
\cite{cccbdb}.
In the top panel of Figure~\ref{fig_lih_pot_compare} we demonstrate the results of this classical calculation as a benchmark and
fit the PES with 
a harmonic, Morse, and cubic spline (without smoothing).
To achieve a reasonable fit for the harmonic potential only five points on either side of the equilibrium geometry were used for the fit, which produces the most accurate vibrational levels and thermodynamic estimates possible for that potential. The harmonic potential, of course, only has reasonable fit in the vicinity of the equilibrium geometry where the bond-stretching potential is nearly parabolic. The Morse potential fit provides good results,
but under-predicts the energy in the region around 3.5 \AA{} and over-predicts the energy around 4.5 \AA.
By definition, the cubic-spline fits the computed PES perfectly at the chosen points.

From the fits of the PES, eigenvalues were computed analytically for the harmonic and Morse potentials, and numerically for the spline.
The eigenvalues of the PES correspond to vibrational energy levels, which were shifted by their zero-point-energy (the zeroth vibrational level), as show in the middle panel of Figure~\ref{fig_lih_pot_compare}.

The bottom panel of Figure~\ref{fig_lih_pot_compare} shows the constant pressure heat capacity that has been computed from the partition function using the vibrational energy levels from the methods described above.
Since the numerical eigenvalues from the cubic spline are very likely the most accurate representation of the eigenvalues of the true underlying PES, we will use these as our reference case.
Note that $C_p$ computed from the cubic spline fit is also the best match to experimental data from NIST \cite{nist_data}.

We should notice that the harmonic levels are evenly spaced,
which is not realistic.
As a result, energy levels of the harmonic potential quickly diverge from those of the cubic spline.
In addition, the harmonic potential fit results in inaccurate prediction of $C_p$, which diverges from the experimental data substantially as temperature increases.
This is likely a result of the evenly spaced energy levels, for which the error becomes greater as the energy (temperature) increases.
Although the Morse potential exhibits energy levels with the correct decrease in spacing as a function of energy, they also diverge from the cubic spline at higher energy levels.
$C_p$ computed from the Morse potential is much better than the harmonic potential, but still diverges from the experimental and spline based calculation as temperature increases.
The prediction of $C_p$ from the cubic spline is very accurate, when compared to experimental data.
Although it too diverges at high temperature, the cubic spline is significantly better than the other fits of the PES.
In this case, the divergence of $C_p$ from experimental data is likely due to limitations in the methods of statistical mechanics used to estimate thermodynamic properties, rather than inaccurate vibrational energy levels.%

Possible sources of high temperature error from the statistical mechanical methods are
that as temperature increases the forces generated from the rotational speed of the molecule begin to stretch the bond between Li and H \textemdash a phenomena called centrifugal distortion.
Corrections to this phenomena, and similar effects of temperature extremes, are beyond the scope of this work, but Ref.~\citenum{popovas_2016} describes methods to perform these corrections, and their influence on prediction thermodynamic properties.
It should also be noted that our treatment of the partition function assumes that the contributions (translation, rotation, vibration, etc.) are separable, which may not be true for all temperatures.
In addition, as temperature increases excited electronic states will become accessible, which makes our assumption that $q_{electronic}=1$ invalid.

Nonetheless, the accuracy of the spline fit compared to other methods underpins our assertion that large improvements are possible in prediction of thermodynamic properties at minimal additional computational cost (comparing the numerical eigenvalues of the cubic spline to the harmonic or Morse potential). 
Regarding the accuracy of our calculations, the cubic spline is a much more flexible method that will enable its use for bending modes (where the Morse potential will not work), and perhaps rotational modes (where neither the Morse or harmonic potentials will work).

\FloatBarrier
\subsection{Experimental results on a quantum computer}
\label{sec_quantum_computer}
\begin{figure}
    \centering    
    \begin{subfigure}[t]{0.4\textwidth}
        \includegraphics[width=1\textwidth]{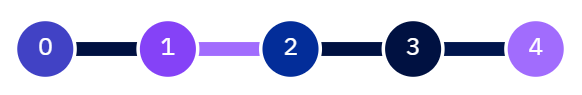}
    \end{subfigure}%
    
    \begin{subfigure}[t]{\textwidth}
		\centering
		\scalebox{0.58}{
		\begin{quantikz}
		\lstick{$q_1: \ket{0}$} 
            &\gate{R_Y(\theta_1)}    &\ctrl{1} &\qw      &\qw
            &\gate{R_Y(\theta_5)}    &\ctrl{1} &\qw      &\qw
            &\gate{R_Y(\theta_9)}    &\ctrl{1} &\qw      &\qw
            &\gate{R_Y(\theta_{13})} &\ctrl{1} &\qw      &\qw
            &\gate{R_Y(\theta_{17})} &\gate{X} &\qw\\
		\lstick{$q_2: \ket{0}$} 
            &\gate{R_Y(\theta_2)}    &\gate{X} &\ctrl{1} &\qw
            &\gate{R_Y(\theta_6)}    &\gate{X} &\ctrl{1} &\qw
            &\gate{R_Y(\theta_{10})} &\gate{X} &\ctrl{1} &\qw
            &\gate{R_Y(\theta_{14})} &\gate{X} &\ctrl{1} &\qw
            &\gate{R_Y(\theta_{18})} &\gate{X} &\qw\\
		\lstick{$q_3: \ket{0}$}
            &\gate{R_Y(\theta_3)}    &\qw      &\gate{X} &\ctrl{1}
            &\gate{R_Y(\theta_7)}    &\qw      &\gate{X} &\ctrl{1}
            &\gate{R_Y(\theta_{11})} &\qw      &\gate{X} &\ctrl{1}
            &\gate{R_Y(\theta_{15})} &\qw      &\gate{X} &\ctrl{1}
            &\gate{R_Y(\theta_{19})} &\qw      &\qw\\
		\lstick{$q_4: \ket{0}$}
            &\gate{R_Y(\theta_4)}    &\qw      &\qw      &\gate{X}
            &\gate{R_Y(\theta_8)}    &\qw      &\qw      &\gate{X}
            &\gate{R_Y(\theta_{12})} &\qw      &\qw      &\gate{X}
            &\gate{R_Y(\theta_{16})} &\qw      &\qw      &\gate{X}          
            &\gate{R_Y(\theta_{20})} &\qw      &\qw
		\end{quantikz}
		}
    \end{subfigure}

    \caption{\label{fig_lih_circuit_rome} Schematic diagram of the topology of ibmq$\_$rome (top) and the quantum circuit used to for VQE (bottom).}
\end{figure}

Following the results described in the previous sections, we implemented our work-flow on a superconducting transmon quantum device with 5-qubits, called \textit{ibmq\_rome}.
We chose four ansatz repetitions as shown in Figure~\ref{fig_lih_circuit_rome}, and 512 measurement shots to evaluate its expectation value at each iteration of the VQE algorithm.
Although 64 shots seem to be effective in simulations, we found that for real hardware 512 shots were required to achieve reasonable accuracy due to additional sources of noise.
For the optimizer, a tolerance for the convergence of the objective function of $10^{-3}$ Ha was used.
Even though resampling was enabled, the standard deviation threshold of $10^{-2}$ Ha required to activate resampling was not reached, so resampling did not occur during these experiments.

\begin{figure}
    \centering    

    \includegraphics[width=0.6\textwidth]{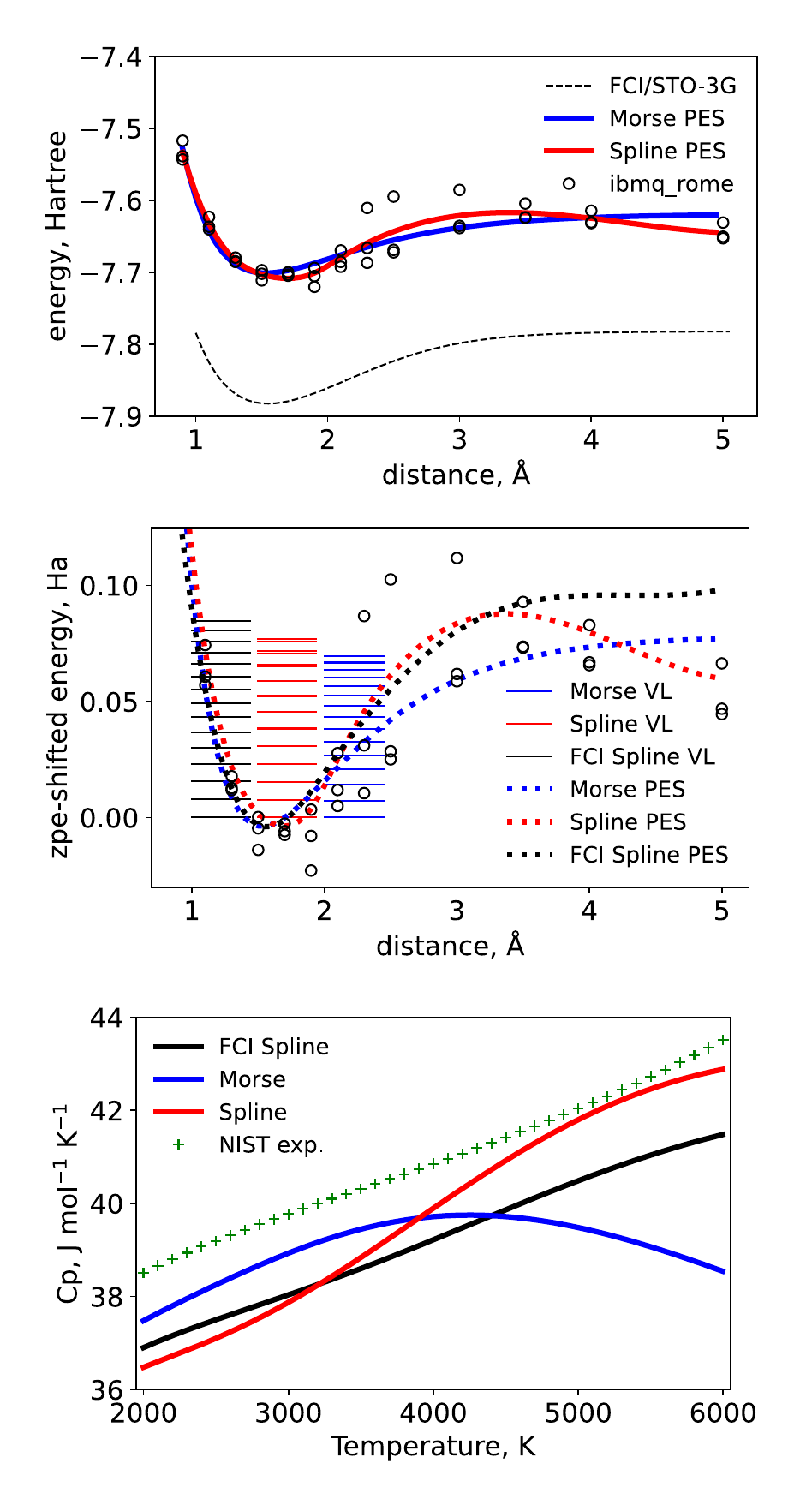}

    \caption{\label{fig_lih_rome} 
    Results from \textit{ibmq\_rome} for LiH.  The PES, vibrational energy levels, and $C_p$ are shown in the top, middle, and bottom panels, respectively.
    Green ``plus signs'' are experimental data.
    The spline fits of the \textit{ibmq\_rome} data were smoothed to enable a better fit to
    the sparse data.}
\end{figure}

Figure~\ref{fig_lih_rome} shows the results of the full work-flow based on hardware experiments with ibmq$\_$rome.
The top panel shows the PES constructed using fewer points than used in the simulations.
Although there is a positive energy offset compared to the FCI/STO-3G reference calculation due to hardware noise, this offset does not affect the resulting vibrational energy levels.

The center panel shows the vibrational energy levels computed by fitting the data to the Morse potential and cubic spline. These values are compared to the FCI/STO-3G energies cubic spline.
Due to the sparsity of data, fitting a harmonic potential was problematic, so it was omitted from this comparison.

Results show that the cubic spline based values match the baseline FCI spline vibrational energy levels better than the Morse potential. Nonetheless, if the data were very noisy, or more sparse, the Morse potential may prove useful since its functional form constrains the shape of the curve to mimic bond stretching, which would be more robust when fitting noisy data.

Finally the lower panel of Figure~\ref{fig_lih_rome}, demonstrates the constant pressure heat capacity. 
The vibrational energy levels from the experiment on \textit{ibmq\_rome} were able to produce a reasonable prediction of $C_p$, compared to the FCI/STO-3G spline fit, and experimental data.
We note that although the spline-based prediction appears closer to experimental data, this is a coincidence, since if the calculation on \textit{ibmq\_rome} was error free, it would match the FCI/STO-3G reference calculation exactly.

\FloatBarrier
\section{Conclusions}

In this work we introduced a robust statistical procedure to evaluate ground state energies and estimate the PES of molecular systems.
Our approach is dictated from the need to create a work-flow for more accurate calculations of thermodynamic observables. 
Based on this we use the accuracy in the calculation of thermodynamic quantities as a metric for the evaluation of hybrid quantum/classical schemes.  

For the calculation of the PES, we highlight that plain implementation of the VQE can be restrictive and advanced schemes can provide significant benefit. 
Namely, resampling protocols improve the reproducibility of the PES, and bootstrapping enables convergence to the minimum eigenvalue by providing better initial conditions at each point along the PES.
In addition, these procedures increase the efficiency of VQE and
allow construction of the PES using fewer measurement shots.
We notice that for the starting point in the PES the equilibrium geometry is a good starting point,
since it exhibits low electron correlation and is represented well by the HF state.

Determining the vibrational energy levels from the PES is another important aspect of the work-flow to calculate thermodynamic observables. 
We noticed substantial improvements 
in the predicted thermodynamic properties when we use
an arbitrary functional form (cubic spline) and numerical eigensolver, compared to the commonly used harmonic or Morse potential and their analytical eigenvalues.
While this approach is limiting 
for bigger molecules, where the decoupling of modes will be needed to reduce the dimensionality of PES, it serves as an example of the kind of improvements that would be needed to fully take advantage of the accurate electronic energies available from quantum computing algorithms.

Beyond the simulation of the work-flow we execute experiments on state-of-the-art IBM quantum hardware (ibmq$\_$rome) that highlight the need of optimizing the work-flow and show good agreement with the findings of our simulations. 
In conclusion, we show that using our work-flow we are 
capable of predicting thermodynamic properties with high accuracy compared to reference calculations and experimental data.

\begin{acknowledgements}
The authors thank Pauline J. Ollitrault for insights and discussions on vibrational structure calculations, and Sumathy Raman for
advice on theoretical aspects of this work.
In addition, we thank Laurent White and Gilad Ben-Shach for their comments and suggestions.
\end{acknowledgements}

\appendix

\section{VQE Classical Optimizer}
\label{app:VQE_classical_opt}
For  SGD is an optimization method defined by the iteration
\[
\btheta^{k+1} = \btheta^k - \gamma^k \mathbf{g}^k.
\]
Here at iteration $k$, $\gamma^k$ is the step length, 
and $\mathbf{g}^k$ is the estimate of the gradient of the objective function, i.e. the expectation value $f = E_\mathcal{H}(\psi(\cdot))$, at $\btheta^k$.
Gradient estimates are obtained using the parameter shift rule \cite{schuld2019evaluating,crooks_2019}
(other techniques for estimating gradients are discussed in Ref.~\citenum{mitarai2020theory}).
These estimates are noisy, but unbiased
(which is why we refer to this method specifically as ``stochastic'' gradient descent).
We use a step length $\gamma^k$ equal to $1$ for the first ten iterations, then decrease it by setting it to $\sfrac{1}{(k-10)}$, for $k > 10$;
this is consistent with most convergence theories for SGD \cite{bertsekas_2000}.
We use a termination criterion based on the change in windowed averages of the objective function and ansatz parameter values.
Specifically, the candidate ansatz parameter values $\btheta^k$ and objective estimates $\hat{f}^k$ from the last $10$ iterations are saved and the method terminates when we have either
\[
    \frac{1}{10}\abs{\hat{f}^{k} - \hat{f}^{k-10}} \le \varepsilon_f
    \qquad \text{or} \qquad
    \frac{1}{10}\norm{ \btheta^{k} - \btheta^{k-10}} \le \varepsilon_{\theta},
\]
for given tolerances $\varepsilon_f$ and $\varepsilon_{\theta}$.

\section{Calculation of Vibrational Energy Levels}
\label{app:vibrational}

Vibrational energies for the harmonic and Morse potential can be expressed analytically.

The harmonic potential is given as
\begin{equation}
V_{h}(r) = \sfrac{1}{2}kr^2
\label{eq_harm}
\end{equation}
where $V_{h}(r)$ is the potential energy,
$r$ is the bond length, and the parameter is $k$, analogous to a spring constant.
The fundamental frequency of this potential is
\begin{equation}
\omega_{0} =   \frac{1}{2 \pi} \; \sqrt{\frac{k}{m_R}},
\end{equation}
where $m_R$ is the reduced mass.
The harmonic potential has analytical eigenvalues
\begin{equation}
\epsilon_{n} = h \omega_{0} (n + \sfrac{1}{2}),
\quad n=0,1,2,\dots,
\end{equation}
where $h$ is Planck's constant.

The Morse potential is   
\begin{equation}
V_{M}(r) = D_e(1 - e^{-\alpha (r-r_0)})^2,
\label{eq_morse}
\end{equation}
where $V_{M}(r)$ is the potential energy
\cite{morse_1929}.  
The parameters are potential energy well depth, $D_e$, the width of the
potential energy well, $\alpha$, which is related to the 
force constant between the atoms, and the equilibrium
bond length $r_0$.
The eigenvalues $\epsilon_n$ of the Morse potential are
\begin{equation}
\displaystyle \epsilon_{n}=h \nu _{0}(n+1/2)-
\frac{\left[h \nu _{0}(n+1/2)\right]^{2}}{4D_{e}},
\quad n=0,1,2, \dots, n_{max},
\label{eq_morse_eigen}
\end{equation}
where
$\nu _{0}$ is the fundamental vibrational frequency, which is related to the parameters of the Morse potential by
\begin{equation}
\nu _{0}= \frac{\alpha}{2\pi } \sqrt \frac{2D_{e}}{m_R}.
\label{eq_morse_fund}
\end{equation}

Finally, numerical techniques for partial differential equations were used to compute the vibrational energies, without confining the PES to a specified analytic function. This was done by using cubic splines to represent the PES, discretizing \cite{larsson2008partial}
the Schr{\"o}dinger PDE using a finite difference scheme, and numerically computing the eigenvalues of the resulting discrete operator.
We used the \texttt{splrep} function in \texttt{SciPy} for the spline 
and \texttt{eigvalsh\_tridiagonal} for computing the eigenvalues \cite{2020SciPy-NMeth}.

\section{Calculation of the partition function}
\label{app:partition_function}
After evaluating the vibrational energies, the next step is to calculate the partition function.
The calculation of the individual contributions
to the molecular partition function are described 
in detail in Ref.~\citenum{book_mcquarrie} and details pertinent to our analysis are highlighted here..
The translational partition function can be calculated classically as
\begin{equation}
q_{translation} = 
\left(
\frac{2 \: \pi \: m_{R} \: k_B \: T}
{h^2}
\right)^{3/2} \: V,
\end{equation}
where volume, $V$, is obtained from the ideal gas law,
$V = k_B \, T / P$.

The rotational partition function for
a heteronuclear diatomic molecule, is calculated as
\begin{equation}
q_{rotational}^{equilib}(T) =
\displaystyle\sum_{J=0}^{J_{max}} (2J + 1) e^{-\Theta_r J(J+1)/T}
\label{eq_qrot_hetnuc}
\end{equation}
where $J$ is the rotational mode (an integer),
and the rotational temperature, $\Theta_r$, is
\begin{equation}
\Theta_r  = \frac{\hslash^2}{2 I_M k_B},
\end{equation}
where 
$I_M$ is the moment of inertia at the equilibrium
bond length.
The energy of each rotational mode, $\epsilon_{rot}$, is
\begin{equation}
\epsilon_{rot} = \frac{\hslash^2}{2 I_M} J(J+1).
\end{equation}

The vibrational partition function is computed from the
Boltzmann summation of the vibrational energy levels
from the PES:
\begin{equation}
q_{vibrational} =
\displaystyle\sum_{n} ^{n_{max}}
e^{-\epsilon_n / k_B \, T}.
\label{eq_qvib}
\end{equation}

\section{Statistical analysis of the PES and thermodynamic predictions}
\label{sec_procedure_mae}

To evaluate the accuracy of VQE calculations 
and predicted thermodynamic properties
we performed simulations of the quantum
computer (with only shot noise included) and compared the results
to the exact calculations (FCI/STO-3G) for the LiH system.
The procedure used is described here.
\begin{enumerate}
    \item The PES was constructed using VQE algorithm as
    described in the main text using the SGD+BR classical optimizer. 
Ground state energy of LiH at interatomic distances from 1 to 5 \AA{} with intervals of 0.1 \AA{} were computed. The energy at the equilibrium geometry (about 1.5 \AA) was evaluated first. Optimized parameters from this geometry were used as initial parameters for geometries proceeding outward toward longer and shorter bond lengths and the process was continued.
    This process of PES generation was repeated 15 times overall, in 5 batches of 3 repeats. 
    \item  A cubic spline was then fit to the mean value of the three repeats from each batch obtained for every geometry. This yielded five determinations of the spline fit for the PES. Averaging over the three repeats in each batch helped mitigate the noise present in the PES construction.
    \item  Using the spline fit, a numerical eigensolver was used to compute the vibrational energy levels for each of the five PES, as discussed in Section \ref{sec_computing_eigvls}.
    \item From the vibrational energy levels and subsequently, the  partition function, constant pressure heat capacity ($C_p$) was then calculated as described in Section \ref{sec:thermo_observables}. This yielded five independent determinations of $C_p$, from which statistics were calculated.               
\end{enumerate}
A corresponding reference calculation was also performed with Psi4 \cite{psi4} using intervals of 0.05 \AA{} and the same basis set (i.e., FCI/STO-3G). The procedure to compute vibrational levels and $C_p$ was similarly also applied to the reference calculation.

From these calculations, the mean absolute error (MAE) was computed for the PES as follows:
\begin{equation}
MAE_{PES} = \displaystyle\frac{1}{n}
\displaystyle\sum\abs{PES_{VQE} - PES_{FCI/STO-3G}}
\end{equation}
where $n$ is the number of points in the PES.
$PES_{VQE}$ and $PES_{FCI/STO-3G}$ refer to the PES as computed by VQE and the reference calculation, respectively. The reported MAE was calculated over the 5 replicates of the PES. For $C_p$, the MAE was computed over the temperature range 2000 - 6000 K, at an interval of 5 K from the five independent replicates. This corresponds to the temperature range for which gas-phase heat capacity data from NIST was available for comparison. 


\begin{thebibliography}{49}%
\makeatletter
\providecommand \@ifxundefined [1]{%
 \@ifx{#1\undefined}
}%
\providecommand \@ifnum [1]{%
 \ifnum #1\expandafter \@firstoftwo
 \else \expandafter \@secondoftwo
 \fi
}%
\providecommand \@ifx [1]{%
 \ifx #1\expandafter \@firstoftwo
 \else \expandafter \@secondoftwo
 \fi
}%
\providecommand \natexlab [1]{#1}%
\providecommand \enquote  [1]{``#1''}%
\providecommand \bibnamefont  [1]{#1}%
\providecommand \bibfnamefont [1]{#1}%
\providecommand \citenamefont [1]{#1}%
\providecommand \href@noop [0]{\@secondoftwo}%
\providecommand \href [0]{\begingroup \@sanitize@url \@href}%
\providecommand \@href[1]{\@@startlink{#1}\@@href}%
\providecommand \@@href[1]{\endgroup#1\@@endlink}%
\providecommand \@sanitize@url [0]{\catcode `\\12\catcode `\$12\catcode
  `\&12\catcode `\#12\catcode `\^12\catcode `\_12\catcode `\%12\relax}%
\providecommand \@@startlink[1]{}%
\providecommand \@@endlink[0]{}%
\providecommand \url  [0]{\begingroup\@sanitize@url \@url }%
\providecommand \@url [1]{\endgroup\@href {#1}{\urlprefix }}%
\providecommand \urlprefix  [0]{URL }%
\providecommand \Eprint [0]{\href }%
\providecommand \doibase [0]{https://doi.org/}%
\providecommand \selectlanguage [0]{\@gobble}%
\providecommand \bibinfo  [0]{\@secondoftwo}%
\providecommand \bibfield  [0]{\@secondoftwo}%
\providecommand \translation [1]{[#1]}%
\providecommand \BibitemOpen [0]{}%
\providecommand \bibitemStop [0]{}%
\providecommand \bibitemNoStop [0]{.\EOS\space}%
\providecommand \EOS [0]{\spacefactor3000\relax}%
\providecommand \BibitemShut  [1]{\csname bibitem#1\endcsname}%
\let\auto@bib@innerbib\@empty
\bibitem [{\citenamefont {Kontogeorgis}\ \emph {et~al.}(2021)\citenamefont
  {Kontogeorgis}, \citenamefont {Dohrn}, \citenamefont {Economou},
  \citenamefont {de~Hemptinne}, \citenamefont {ten Kate}, \citenamefont
  {Kuitunen}, \citenamefont {Mooijer}, \citenamefont {Zilnik},\ and\
  \citenamefont {Vesovic}}]{kontogeorgis_2021}%
  \BibitemOpen
  \bibfield  {author} {\bibinfo {author} {\bibfnamefont {G.~M.}\ \bibnamefont
  {Kontogeorgis}}, \bibinfo {author} {\bibfnamefont {R.}~\bibnamefont {Dohrn}},
  \bibinfo {author} {\bibfnamefont {I.~G.}\ \bibnamefont {Economou}}, \bibinfo
  {author} {\bibfnamefont {J.-C.}\ \bibnamefont {de~Hemptinne}}, \bibinfo
  {author} {\bibfnamefont {A.}~\bibnamefont {ten Kate}}, \bibinfo {author}
  {\bibfnamefont {S.}~\bibnamefont {Kuitunen}}, \bibinfo {author}
  {\bibfnamefont {M.}~\bibnamefont {Mooijer}}, \bibinfo {author} {\bibfnamefont
  {L.~F.}\ \bibnamefont {Zilnik}},\ and\ \bibinfo {author} {\bibfnamefont
  {V.}~\bibnamefont {Vesovic}},\ }\bibfield  {title} {\bibinfo {title}
  {Industrial requirements for thermodynamic and transport properties: 2020},\
  }\bibfield  {journal} {\bibinfo  {journal} {Industrial and Engineering
  Chemistry Research}\ }\href {https://doi.org/10.1021/acs.iecr.0c05356}
  {10.1021/acs.iecr.0c05356} (\bibinfo {year} {2021})\BibitemShut {NoStop}%
\bibitem [{\citenamefont {Green}(2020)}]{green_2020}%
  \BibitemOpen
  \bibfield  {author} {\bibinfo {author} {\bibfnamefont {W.~H.}\ \bibnamefont
  {Green}},\ }\bibfield  {title} {\bibinfo {title} {Moving from postdictive to
  predictive kinetics in reaction engineering},\ }\href {https://doi.org/ARTN
  e17059 10.1002/aic.17059} {\bibfield  {journal} {\bibinfo  {journal} {AIChE
  Journal}\ }\textbf {\bibinfo {volume} {66}},\ \bibinfo {pages} {e17059}
  (\bibinfo {year} {2020})}\BibitemShut {NoStop}%
\bibitem [{\citenamefont {Abrams}\ and\ \citenamefont
  {Lloyd}(1997)}]{abrams_1997}%
  \BibitemOpen
  \bibfield  {author} {\bibinfo {author} {\bibfnamefont {D.~S.}\ \bibnamefont
  {Abrams}}\ and\ \bibinfo {author} {\bibfnamefont {S.}~\bibnamefont {Lloyd}},\
  }\bibfield  {title} {\bibinfo {title} {Simulation of many-body fermi systems
  on a universal quantum computer},\ }\href
  {https://doi.org/10.1103/PhysRevLett.79.2586} {\bibfield  {journal} {\bibinfo
   {journal} {Physical Review Letters}\ }\textbf {\bibinfo {volume} {79}},\
  \bibinfo {pages} {2586} (\bibinfo {year} {1997})}\BibitemShut {NoStop}%
\bibitem [{\citenamefont {Kassal}\ \emph {et~al.}(2011)\citenamefont {Kassal},
  \citenamefont {Whitfield}, \citenamefont {Perdomo-Ortiz}, \citenamefont
  {Yung},\ and\ \citenamefont {Aspuru-Guzik}}]{kassal_2011}%
  \BibitemOpen
  \bibfield  {author} {\bibinfo {author} {\bibfnamefont {I.}~\bibnamefont
  {Kassal}}, \bibinfo {author} {\bibfnamefont {J.~D.}\ \bibnamefont
  {Whitfield}}, \bibinfo {author} {\bibfnamefont {A.}~\bibnamefont
  {Perdomo-Ortiz}}, \bibinfo {author} {\bibfnamefont {M.-H.}\ \bibnamefont
  {Yung}},\ and\ \bibinfo {author} {\bibfnamefont {A.}~\bibnamefont
  {Aspuru-Guzik}},\ }\bibfield  {title} {\bibinfo {title} {Simulating chemistry
  using quantum computers},\ }\href
  {https://doi.org/10.1146/annurev-physchem-032210-103512} {\bibfield
  {journal} {\bibinfo  {journal} {Annual Review of Physical Chemistry}\
  }\textbf {\bibinfo {volume} {62}},\ \bibinfo {pages} {185} (\bibinfo {year}
  {2011})}\BibitemShut {NoStop}%
\bibitem [{\citenamefont {O'Malley}\ \emph {et~al.}(2016)\citenamefont
  {O'Malley}, \citenamefont {Babbush}, \citenamefont {Kivlichan}, \citenamefont
  {Romero}, \citenamefont {McClean}, \citenamefont {Barends}, \citenamefont
  {Kelly}, \citenamefont {Roushan}, \citenamefont {Tranter}, \citenamefont
  {Ding}, \citenamefont {Campbell}, \citenamefont {Chen}, \citenamefont {Chen},
  \citenamefont {Chiaro}, \citenamefont {Dunsworth}, \citenamefont {Fowler},
  \citenamefont {Jeffrey}, \citenamefont {Lucero}, \citenamefont {Megrant},
  \citenamefont {Mutus}, \citenamefont {Neeley}, \citenamefont {Neill},
  \citenamefont {Quintana}, \citenamefont {Sank}, \citenamefont {Vainsencher},
  \citenamefont {Wenner}, \citenamefont {White}, \citenamefont {Coveney},
  \citenamefont {Love}, \citenamefont {Neven}, \citenamefont {Aspuru-Guzik},\
  and\ \citenamefont {Martinis}}]{omalley_2016}%
  \BibitemOpen
  \bibfield  {author} {\bibinfo {author} {\bibfnamefont {P.~J.}\ \bibnamefont
  {O'Malley}}, \bibinfo {author} {\bibfnamefont {R.}~\bibnamefont {Babbush}},
  \bibinfo {author} {\bibfnamefont {I.~D.}\ \bibnamefont {Kivlichan}}, \bibinfo
  {author} {\bibfnamefont {J.}~\bibnamefont {Romero}}, \bibinfo {author}
  {\bibfnamefont {J.~R.}\ \bibnamefont {McClean}}, \bibinfo {author}
  {\bibfnamefont {R.}~\bibnamefont {Barends}}, \bibinfo {author} {\bibfnamefont
  {J.}~\bibnamefont {Kelly}}, \bibinfo {author} {\bibfnamefont
  {P.}~\bibnamefont {Roushan}}, \bibinfo {author} {\bibfnamefont
  {A.}~\bibnamefont {Tranter}}, \bibinfo {author} {\bibfnamefont
  {N.}~\bibnamefont {Ding}}, \bibinfo {author} {\bibfnamefont {B.}~\bibnamefont
  {Campbell}}, \bibinfo {author} {\bibfnamefont {Y.}~\bibnamefont {Chen}},
  \bibinfo {author} {\bibfnamefont {Z.}~\bibnamefont {Chen}}, \bibinfo {author}
  {\bibfnamefont {B.}~\bibnamefont {Chiaro}}, \bibinfo {author} {\bibfnamefont
  {A.}~\bibnamefont {Dunsworth}}, \bibinfo {author} {\bibfnamefont {A.~G.}\
  \bibnamefont {Fowler}}, \bibinfo {author} {\bibfnamefont {E.}~\bibnamefont
  {Jeffrey}}, \bibinfo {author} {\bibfnamefont {E.}~\bibnamefont {Lucero}},
  \bibinfo {author} {\bibfnamefont {A.}~\bibnamefont {Megrant}}, \bibinfo
  {author} {\bibfnamefont {J.~Y.}\ \bibnamefont {Mutus}}, \bibinfo {author}
  {\bibfnamefont {M.}~\bibnamefont {Neeley}}, \bibinfo {author} {\bibfnamefont
  {C.}~\bibnamefont {Neill}}, \bibinfo {author} {\bibfnamefont
  {C.}~\bibnamefont {Quintana}}, \bibinfo {author} {\bibfnamefont
  {D.}~\bibnamefont {Sank}}, \bibinfo {author} {\bibfnamefont {A.}~\bibnamefont
  {Vainsencher}}, \bibinfo {author} {\bibfnamefont {J.}~\bibnamefont {Wenner}},
  \bibinfo {author} {\bibfnamefont {T.~C.}\ \bibnamefont {White}}, \bibinfo
  {author} {\bibfnamefont {P.~V.}\ \bibnamefont {Coveney}}, \bibinfo {author}
  {\bibfnamefont {P.~J.}\ \bibnamefont {Love}}, \bibinfo {author}
  {\bibfnamefont {H.}~\bibnamefont {Neven}}, \bibinfo {author} {\bibfnamefont
  {A.}~\bibnamefont {Aspuru-Guzik}},\ and\ \bibinfo {author} {\bibfnamefont
  {J.~M.}\ \bibnamefont {Martinis}},\ }\bibfield  {title} {\bibinfo {title}
  {Scalable quantum simulation of molecular energies},\ }\href@noop {}
  {\bibfield  {journal} {\bibinfo  {journal} {Physical Review X}\ }\textbf
  {\bibinfo {volume} {6}},\ \bibinfo {pages} {031007} (\bibinfo {year}
  {2016})}\BibitemShut {NoStop}%
\bibitem [{\citenamefont {Kandala}\ \emph {et~al.}(2017)\citenamefont
  {Kandala}, \citenamefont {Mezzacapo}, \citenamefont {Temme}, \citenamefont
  {Takita}, \citenamefont {Brink}, \citenamefont {Chow},\ and\ \citenamefont
  {Gambetta}}]{kandala_2017}%
  \BibitemOpen
  \bibfield  {author} {\bibinfo {author} {\bibfnamefont {A.}~\bibnamefont
  {Kandala}}, \bibinfo {author} {\bibfnamefont {A.}~\bibnamefont {Mezzacapo}},
  \bibinfo {author} {\bibfnamefont {K.}~\bibnamefont {Temme}}, \bibinfo
  {author} {\bibfnamefont {M.}~\bibnamefont {Takita}}, \bibinfo {author}
  {\bibfnamefont {M.}~\bibnamefont {Brink}}, \bibinfo {author} {\bibfnamefont
  {J.~M.}\ \bibnamefont {Chow}},\ and\ \bibinfo {author} {\bibfnamefont
  {J.~M.}\ \bibnamefont {Gambetta}},\ }\bibfield  {title} {\bibinfo {title}
  {Hardware-efficient variational quantum eigensolver for small molecules and
  quantum magnets},\ }\href@noop {} {\bibfield  {journal} {\bibinfo  {journal}
  {Nature}\ }\textbf {\bibinfo {volume} {549}},\ \bibinfo {pages} {242}
  (\bibinfo {year} {2017})}\BibitemShut {NoStop}%
\bibitem [{\citenamefont {Barkoutsos}\ \emph {et~al.}(2018)\citenamefont
  {Barkoutsos}, \citenamefont {Gonthier}, \citenamefont {Sokolov},
  \citenamefont {Moll}, \citenamefont {Salis}, \citenamefont {Fuhrer},
  \citenamefont {Ganzhorn}, \citenamefont {Egger}, \citenamefont {Troyer},
  \citenamefont {Mezzacapo}, \citenamefont {Filipp},\ and\ \citenamefont
  {Tavernelli}}]{barkoutsos_2018}%
  \BibitemOpen
  \bibfield  {author} {\bibinfo {author} {\bibfnamefont {P.~K.}\ \bibnamefont
  {Barkoutsos}}, \bibinfo {author} {\bibfnamefont {J.~F.}\ \bibnamefont
  {Gonthier}}, \bibinfo {author} {\bibfnamefont {I.}~\bibnamefont {Sokolov}},
  \bibinfo {author} {\bibfnamefont {N.}~\bibnamefont {Moll}}, \bibinfo {author}
  {\bibfnamefont {G.}~\bibnamefont {Salis}}, \bibinfo {author} {\bibfnamefont
  {A.}~\bibnamefont {Fuhrer}}, \bibinfo {author} {\bibfnamefont
  {M.}~\bibnamefont {Ganzhorn}}, \bibinfo {author} {\bibfnamefont {D.~J.}\
  \bibnamefont {Egger}}, \bibinfo {author} {\bibfnamefont {M.}~\bibnamefont
  {Troyer}}, \bibinfo {author} {\bibfnamefont {A.}~\bibnamefont {Mezzacapo}},
  \bibinfo {author} {\bibfnamefont {S.}~\bibnamefont {Filipp}},\ and\ \bibinfo
  {author} {\bibfnamefont {I.}~\bibnamefont {Tavernelli}},\ }\bibfield  {title}
  {\bibinfo {title} {Quantum algorithms for electronic structure calculations:
  Particle-hole hamiltonian and optimized wave-function expansions},\
  }\href@noop {} {\bibfield  {journal} {\bibinfo  {journal} {Physical Review
  A}\ }\textbf {\bibinfo {volume} {98}},\ \bibinfo {pages} {022322} (\bibinfo
  {year} {2018})}\BibitemShut {NoStop}%
\bibitem [{\citenamefont {Hempel}\ \emph {et~al.}(2018)\citenamefont {Hempel},
  \citenamefont {Maier}, \citenamefont {Romero}, \citenamefont {McClean},
  \citenamefont {Monz}, \citenamefont {Shen}, \citenamefont {Jurcevic},
  \citenamefont {Lanyon}, \citenamefont {Love}, \citenamefont {Babbush},
  \citenamefont {Aspuru-Guzik}, \citenamefont {Blatt},\ and\ \citenamefont
  {Roos}}]{hempel_2018}%
  \BibitemOpen
  \bibfield  {author} {\bibinfo {author} {\bibfnamefont {C.}~\bibnamefont
  {Hempel}}, \bibinfo {author} {\bibfnamefont {C.}~\bibnamefont {Maier}},
  \bibinfo {author} {\bibfnamefont {J.}~\bibnamefont {Romero}}, \bibinfo
  {author} {\bibfnamefont {J.}~\bibnamefont {McClean}}, \bibinfo {author}
  {\bibfnamefont {T.}~\bibnamefont {Monz}}, \bibinfo {author} {\bibfnamefont
  {H.}~\bibnamefont {Shen}}, \bibinfo {author} {\bibfnamefont {P.}~\bibnamefont
  {Jurcevic}}, \bibinfo {author} {\bibfnamefont {B.~P.}\ \bibnamefont
  {Lanyon}}, \bibinfo {author} {\bibfnamefont {P.}~\bibnamefont {Love}},
  \bibinfo {author} {\bibfnamefont {R.}~\bibnamefont {Babbush}}, \bibinfo
  {author} {\bibfnamefont {A.}~\bibnamefont {Aspuru-Guzik}}, \bibinfo {author}
  {\bibfnamefont {R.}~\bibnamefont {Blatt}},\ and\ \bibinfo {author}
  {\bibfnamefont {C.~F.}\ \bibnamefont {Roos}},\ }\bibfield  {title} {\bibinfo
  {title} {Quantum chemistry calculations on a trapped-ion quantum simulator},\
  }\href@noop {} {\bibfield  {journal} {\bibinfo  {journal} {Physical Review
  X}\ }\textbf {\bibinfo {volume} {8}},\ \bibinfo {pages} {031022} (\bibinfo
  {year} {2018})}\BibitemShut {NoStop}%
\bibitem [{\citenamefont {Romero}\ \emph {et~al.}(2018)\citenamefont {Romero},
  \citenamefont {Babbush}, \citenamefont {McClean}, \citenamefont {Hempel},
  \citenamefont {Love},\ and\ \citenamefont {Aspuru-Guzik}}]{romero2018}%
  \BibitemOpen
  \bibfield  {author} {\bibinfo {author} {\bibfnamefont {J.}~\bibnamefont
  {Romero}}, \bibinfo {author} {\bibfnamefont {R.}~\bibnamefont {Babbush}},
  \bibinfo {author} {\bibfnamefont {J.~R.}\ \bibnamefont {McClean}}, \bibinfo
  {author} {\bibfnamefont {C.}~\bibnamefont {Hempel}}, \bibinfo {author}
  {\bibfnamefont {P.~J.}\ \bibnamefont {Love}},\ and\ \bibinfo {author}
  {\bibfnamefont {A.}~\bibnamefont {Aspuru-Guzik}},\ }\bibfield  {title}
  {\bibinfo {title} {Strategies for quantum computing molecular energies using
  the unitary coupled cluster ansatz},\ }\href@noop {} {\bibfield  {journal}
  {\bibinfo  {journal} {Quantum Science and Technology}\ }\textbf {\bibinfo
  {volume} {4}},\ \bibinfo {pages} {014008} (\bibinfo {year}
  {2018})}\BibitemShut {NoStop}%
\bibitem [{\citenamefont {Babbush}\ \emph {et~al.}(2018)\citenamefont
  {Babbush}, \citenamefont {Wiebe}, \citenamefont {McClean}, \citenamefont
  {McClain}, \citenamefont {Neven},\ and\ \citenamefont {Chan}}]{babbush2018}%
  \BibitemOpen
  \bibfield  {author} {\bibinfo {author} {\bibfnamefont {R.}~\bibnamefont
  {Babbush}}, \bibinfo {author} {\bibfnamefont {N.}~\bibnamefont {Wiebe}},
  \bibinfo {author} {\bibfnamefont {J.~R.}\ \bibnamefont {McClean}}, \bibinfo
  {author} {\bibfnamefont {J.}~\bibnamefont {McClain}}, \bibinfo {author}
  {\bibfnamefont {H.}~\bibnamefont {Neven}},\ and\ \bibinfo {author}
  {\bibfnamefont {G.~K.~L.}\ \bibnamefont {Chan}},\ }\bibfield  {title}
  {\bibinfo {title} {Low depth quantum simulation of materials},\ }\href@noop
  {} {\bibfield  {journal} {\bibinfo  {journal} {Physical Review X}\ }\textbf
  {\bibinfo {volume} {8}},\ \bibinfo {pages} {011044} (\bibinfo {year}
  {2018})}\BibitemShut {NoStop}%
\bibitem [{\citenamefont {Cao}\ \emph {et~al.}(2019)\citenamefont {Cao},
  \citenamefont {Romero}, \citenamefont {Olson}, \citenamefont {Degroote},
  \citenamefont {Johnson}, \citenamefont {Kieferová}, \citenamefont
  {Kivlichan}, \citenamefont {Menke}, \citenamefont {Peropadre}, \citenamefont
  {Sawaya}, \citenamefont {Sim}, \citenamefont {Veis},\ and\ \citenamefont
  {Aspuru-Guzik}}]{cao_2019}%
  \BibitemOpen
  \bibfield  {author} {\bibinfo {author} {\bibfnamefont {Y.}~\bibnamefont
  {Cao}}, \bibinfo {author} {\bibfnamefont {J.}~\bibnamefont {Romero}},
  \bibinfo {author} {\bibfnamefont {J.~P.}\ \bibnamefont {Olson}}, \bibinfo
  {author} {\bibfnamefont {M.}~\bibnamefont {Degroote}}, \bibinfo {author}
  {\bibfnamefont {P.~D.}\ \bibnamefont {Johnson}}, \bibinfo {author}
  {\bibfnamefont {M.}~\bibnamefont {Kieferová}}, \bibinfo {author}
  {\bibfnamefont {I.~D.}\ \bibnamefont {Kivlichan}}, \bibinfo {author}
  {\bibfnamefont {T.}~\bibnamefont {Menke}}, \bibinfo {author} {\bibfnamefont
  {B.}~\bibnamefont {Peropadre}}, \bibinfo {author} {\bibfnamefont {N.~P.~D.}\
  \bibnamefont {Sawaya}}, \bibinfo {author} {\bibfnamefont {S.}~\bibnamefont
  {Sim}}, \bibinfo {author} {\bibfnamefont {L.}~\bibnamefont {Veis}},\ and\
  \bibinfo {author} {\bibfnamefont {A.}~\bibnamefont {Aspuru-Guzik}},\
  }\bibfield  {title} {\bibinfo {title} {Quantum chemistry in the age of
  quantum computing},\ }\bibfield  {journal} {\bibinfo  {journal} {Chemical
  Reviews}\ }\href {https://doi.org/10.1021/acs.chemrev.8b00803}
  {10.1021/acs.chemrev.8b00803} (\bibinfo {year} {2019})\BibitemShut {NoStop}%
\bibitem [{\citenamefont {Ollitrault}\ \emph {et~al.}(2020)\citenamefont
  {Ollitrault}, \citenamefont {Baiardi}, \citenamefont {Reiher},\ and\
  \citenamefont {Tavernelli}}]{ollitrault_2020}%
  \BibitemOpen
  \bibfield  {author} {\bibinfo {author} {\bibfnamefont {P.~J.}\ \bibnamefont
  {Ollitrault}}, \bibinfo {author} {\bibfnamefont {A.}~\bibnamefont {Baiardi}},
  \bibinfo {author} {\bibfnamefont {M.}~\bibnamefont {Reiher}},\ and\ \bibinfo
  {author} {\bibfnamefont {I.}~\bibnamefont {Tavernelli}},\ }\bibfield  {title}
  {\bibinfo {title} {Hardware efficient quantum algorithms for vibrational
  structure calculations},\ }\href {https://doi.org/10.1039/D0SC01908A}
  {\bibfield  {journal} {\bibinfo  {journal} {Chemical Science}\ }\textbf
  {\bibinfo {volume} {11}},\ \bibinfo {pages} {6842} (\bibinfo {year}
  {2020})}\BibitemShut {NoStop}%
\bibitem [{\citenamefont {Gao}\ \emph {et~al.}(2021)\citenamefont {Gao},
  \citenamefont {Nakamura}, \citenamefont {Gujarati}, \citenamefont {Jones},
  \citenamefont {Rice}, \citenamefont {Wood}, \citenamefont {Pistoia},
  \citenamefont {Garcia},\ and\ \citenamefont
  {Yamamoto}}]{gao2021computational}%
  \BibitemOpen
  \bibfield  {author} {\bibinfo {author} {\bibfnamefont {Q.}~\bibnamefont
  {Gao}}, \bibinfo {author} {\bibfnamefont {H.}~\bibnamefont {Nakamura}},
  \bibinfo {author} {\bibfnamefont {T.~P.}\ \bibnamefont {Gujarati}}, \bibinfo
  {author} {\bibfnamefont {G.~O.}\ \bibnamefont {Jones}}, \bibinfo {author}
  {\bibfnamefont {J.~E.}\ \bibnamefont {Rice}}, \bibinfo {author}
  {\bibfnamefont {S.~P.}\ \bibnamefont {Wood}}, \bibinfo {author}
  {\bibfnamefont {M.}~\bibnamefont {Pistoia}}, \bibinfo {author} {\bibfnamefont
  {J.~M.}\ \bibnamefont {Garcia}},\ and\ \bibinfo {author} {\bibfnamefont
  {N.}~\bibnamefont {Yamamoto}},\ }\bibfield  {title} {\bibinfo {title}
  {Computational investigations of the lithium superoxide dimer rearrangement
  on noisy quantum devices},\ }\href
  {https://pubs.acs.org/doi/10.1021/acs.jpca.0c09530} {\bibfield  {journal}
  {\bibinfo  {journal} {J. Phys. Chem. A}\ }\textbf {\bibinfo {volume} {125}},\
  \bibinfo {pages} {1827–1836} (\bibinfo {year} {2021})}\BibitemShut
  {NoStop}%
\bibitem [{\citenamefont {Rice}\ \emph {et~al.}(2021)\citenamefont {Rice},
  \citenamefont {Gujarati}, \citenamefont {Motta}, \citenamefont {Takeshita},
  \citenamefont {Lee}, \citenamefont {Latone},\ and\ \citenamefont
  {Garcia}}]{jrice_2021}%
  \BibitemOpen
  \bibfield  {author} {\bibinfo {author} {\bibfnamefont {J.~E.}\ \bibnamefont
  {Rice}}, \bibinfo {author} {\bibfnamefont {T.~P.}\ \bibnamefont {Gujarati}},
  \bibinfo {author} {\bibfnamefont {M.}~\bibnamefont {Motta}}, \bibinfo
  {author} {\bibfnamefont {T.~Y.}\ \bibnamefont {Takeshita}}, \bibinfo {author}
  {\bibfnamefont {E.}~\bibnamefont {Lee}}, \bibinfo {author} {\bibfnamefont
  {J.~A.}\ \bibnamefont {Latone}},\ and\ \bibinfo {author} {\bibfnamefont
  {J.~M.}\ \bibnamefont {Garcia}},\ }\bibfield  {title} {\bibinfo {title}
  {Quantum computation of dominant products in lithium–sulfur batteries},\
  }\href {https://doi.org/10.1063/5.0044068} {\bibfield  {journal} {\bibinfo
  {journal} {J. Chem. Phys.}\ }\textbf {\bibinfo {volume} {154}},\ \bibinfo
  {pages} {134115} (\bibinfo {year} {2021})}\BibitemShut {NoStop}%
\bibitem [{\citenamefont {Abrams}\ and\ \citenamefont
  {Lloyd}(1999)}]{abrams_1999}%
  \BibitemOpen
  \bibfield  {author} {\bibinfo {author} {\bibfnamefont {D.~S.}\ \bibnamefont
  {Abrams}}\ and\ \bibinfo {author} {\bibfnamefont {S.}~\bibnamefont {Lloyd}},\
  }\bibfield  {title} {\bibinfo {title} {Quantum algorithm providing
  exponential speed increase for finding eigenvalues and eigenvectors},\ }\href
  {https://doi.org/10.1103/PhysRevLett.83.5162} {\bibfield  {journal} {\bibinfo
   {journal} {Physical Review Letters}\ }\textbf {\bibinfo {volume} {83}},\
  \bibinfo {pages} {5162} (\bibinfo {year} {1999})}\BibitemShut {NoStop}%
\bibitem [{\citenamefont {Bartlett}\ and\ \citenamefont
  {Musiał}(2007)}]{bartlett_2007}%
  \BibitemOpen
  \bibfield  {author} {\bibinfo {author} {\bibfnamefont {R.~J.}\ \bibnamefont
  {Bartlett}}\ and\ \bibinfo {author} {\bibfnamefont {M.}~\bibnamefont
  {Musiał}},\ }\bibfield  {title} {\bibinfo {title} {Coupled-cluster theory in
  quantum chemistry},\ }\href {https://doi.org/10.1103/RevModPhys.79.291}
  {\bibfield  {journal} {\bibinfo  {journal} {Reviews of Modern Physics}\
  }\textbf {\bibinfo {volume} {79}},\ \bibinfo {pages} {291} (\bibinfo {year}
  {2007})}\BibitemShut {NoStop}%
\bibitem [{\citenamefont {Colless}\ \emph {et~al.}(2018)\citenamefont
  {Colless}, \citenamefont {Ramasesh}, \citenamefont {Dahlen}, \citenamefont
  {Blok}, \citenamefont {Kimchi-Schwartz}, \citenamefont {McClean},
  \citenamefont {Carter}, \citenamefont {de~Jong},\ and\ \citenamefont
  {Siddiqi}}]{colless_2018}%
  \BibitemOpen
  \bibfield  {author} {\bibinfo {author} {\bibfnamefont {J.~I.}\ \bibnamefont
  {Colless}}, \bibinfo {author} {\bibfnamefont {V.~V.}\ \bibnamefont
  {Ramasesh}}, \bibinfo {author} {\bibfnamefont {D.}~\bibnamefont {Dahlen}},
  \bibinfo {author} {\bibfnamefont {M.~S.}\ \bibnamefont {Blok}}, \bibinfo
  {author} {\bibfnamefont {M.~E.}\ \bibnamefont {Kimchi-Schwartz}}, \bibinfo
  {author} {\bibfnamefont {J.~R.}\ \bibnamefont {McClean}}, \bibinfo {author}
  {\bibfnamefont {J.}~\bibnamefont {Carter}}, \bibinfo {author} {\bibfnamefont
  {W.~A.}\ \bibnamefont {de~Jong}},\ and\ \bibinfo {author} {\bibfnamefont
  {I.}~\bibnamefont {Siddiqi}},\ }\bibfield  {title} {\bibinfo {title}
  {Computation of molecular spectra on a quantum processor with an
  error-resilient algorithm},\ }\href@noop {} {\bibfield  {journal} {\bibinfo
  {journal} {Physical Review X}\ }\textbf {\bibinfo {volume} {8}},\ \bibinfo
  {pages} {011021} (\bibinfo {year} {2018})}\BibitemShut {NoStop}%
\bibitem [{\citenamefont {Preskill}(2018)}]{preskill2018quantum}%
  \BibitemOpen
  \bibfield  {author} {\bibinfo {author} {\bibfnamefont {J.}~\bibnamefont
  {Preskill}},\ }\bibfield  {title} {\bibinfo {title} {Quantum computing in the
  nisq era and beyond},\ }\href@noop {} {\bibfield  {journal} {\bibinfo
  {journal} {Quantum}\ }\textbf {\bibinfo {volume} {2}},\ \bibinfo {pages} {79}
  (\bibinfo {year} {2018})}\BibitemShut {NoStop}%
\bibitem [{\citenamefont {Roffe}(2019)}]{roffe:19}%
  \BibitemOpen
  \bibfield  {author} {\bibinfo {author} {\bibfnamefont {J.}~\bibnamefont
  {Roffe}},\ }\bibfield  {title} {\bibinfo {title} {Quantum error correction:
  an introductory guide},\ }\href
  {https://doi.org/10.1080/00107514.2019.1667078} {\bibfield  {journal}
  {\bibinfo  {journal} {Contemporary Physics}\ }\textbf {\bibinfo {volume}
  {60}},\ \bibinfo {pages} {226} (\bibinfo {year} {2019})},\ \Eprint
  {https://arxiv.org/abs/https://doi.org/10.1080/00107514.2019.1667078}
  {https://doi.org/10.1080/00107514.2019.1667078} \BibitemShut {NoStop}%
\bibitem [{\citenamefont {Peruzzo}\ \emph {et~al.}(2014)\citenamefont
  {Peruzzo}, \citenamefont {McClean}, \citenamefont {Shadbolt}, \citenamefont
  {Yung}, \citenamefont {Zhou}, \citenamefont {Love}, \citenamefont
  {Aspuru-Guzik},\ and\ \citenamefont {O’Brien}}]{peruzzo_2014}%
  \BibitemOpen
  \bibfield  {author} {\bibinfo {author} {\bibfnamefont {A.}~\bibnamefont
  {Peruzzo}}, \bibinfo {author} {\bibfnamefont {J.}~\bibnamefont {McClean}},
  \bibinfo {author} {\bibfnamefont {P.}~\bibnamefont {Shadbolt}}, \bibinfo
  {author} {\bibfnamefont {M.-H.}\ \bibnamefont {Yung}}, \bibinfo {author}
  {\bibfnamefont {X.-Q.}\ \bibnamefont {Zhou}}, \bibinfo {author}
  {\bibfnamefont {P.~J.}\ \bibnamefont {Love}}, \bibinfo {author}
  {\bibfnamefont {A.}~\bibnamefont {Aspuru-Guzik}},\ and\ \bibinfo {author}
  {\bibfnamefont {J.~L.}\ \bibnamefont {O’Brien}},\ }\bibfield  {title}
  {\bibinfo {title} {A variational eigenvalue solver on a photonic quantum
  processor},\ }\href@noop {} {\bibfield  {journal} {\bibinfo  {journal}
  {Nature Communications}\ }\textbf {\bibinfo {volume} {5}},\ \bibinfo {pages}
  {4213} (\bibinfo {year} {2014})}\BibitemShut {NoStop}%
\bibitem [{\citenamefont {Aspuru-Guzik}\ \emph {et~al.}(2005)\citenamefont
  {Aspuru-Guzik}, \citenamefont {Dutoi}, \citenamefont {Love},\ and\
  \citenamefont {Head-Gordon}}]{aspuru_guzik_2005}%
  \BibitemOpen
  \bibfield  {author} {\bibinfo {author} {\bibfnamefont {A.}~\bibnamefont
  {Aspuru-Guzik}}, \bibinfo {author} {\bibfnamefont {A.~D.}\ \bibnamefont
  {Dutoi}}, \bibinfo {author} {\bibfnamefont {P.~J.}\ \bibnamefont {Love}},\
  and\ \bibinfo {author} {\bibfnamefont {M.}~\bibnamefont {Head-Gordon}},\
  }\bibfield  {title} {\bibinfo {title} {Simulated quantum computation of
  molecular energies},\ }\href@noop {} {\bibfield  {journal} {\bibinfo
  {journal} {Science}\ }\textbf {\bibinfo {volume} {309}},\ \bibinfo {pages}
  {1704} (\bibinfo {year} {2005})}\BibitemShut {NoStop}%
\bibitem [{\citenamefont {Reiher}\ \emph {et~al.}(2017)\citenamefont {Reiher},
  \citenamefont {Wiebe}, \citenamefont {Svore}, \citenamefont {Wecker},\ and\
  \citenamefont {Troyer}}]{reiher2017elucidating}%
  \BibitemOpen
  \bibfield  {author} {\bibinfo {author} {\bibfnamefont {M.}~\bibnamefont
  {Reiher}}, \bibinfo {author} {\bibfnamefont {N.}~\bibnamefont {Wiebe}},
  \bibinfo {author} {\bibfnamefont {K.~M.}\ \bibnamefont {Svore}}, \bibinfo
  {author} {\bibfnamefont {D.}~\bibnamefont {Wecker}},\ and\ \bibinfo {author}
  {\bibfnamefont {M.}~\bibnamefont {Troyer}},\ }\bibfield  {title} {\bibinfo
  {title} {Elucidating reaction mechanisms on quantum computers},\ }\href@noop
  {} {\bibfield  {journal} {\bibinfo  {journal} {Proceedings of the National
  Academy of Sciences}\ }\textbf {\bibinfo {volume} {114}},\ \bibinfo {pages}
  {7555} (\bibinfo {year} {2017})}\BibitemShut {NoStop}%
\bibitem [{\citenamefont {Jordan}\ and\ \citenamefont
  {Wigner}(1928)}]{jordan-wigner_1928}%
  \BibitemOpen
  \bibfield  {author} {\bibinfo {author} {\bibfnamefont {P.}~\bibnamefont
  {Jordan}}\ and\ \bibinfo {author} {\bibfnamefont {E.}~\bibnamefont
  {Wigner}},\ }\bibfield  {title} {\bibinfo {title} {\"{U}ber das paulische
  \"{A}quivalenzverbot},\ }\href@noop {} {\bibfield  {journal} {\bibinfo
  {journal} {Z. Phys}\ }\textbf {\bibinfo {volume} {47}},\ \bibinfo {pages}
  {631} (\bibinfo {year} {1928})}\BibitemShut {NoStop}%
\bibitem [{\citenamefont {Bravyi}\ and\ \citenamefont
  {Kitaev}(2002)}]{bravyi_2002}%
  \BibitemOpen
  \bibfield  {author} {\bibinfo {author} {\bibfnamefont {S.}~\bibnamefont
  {Bravyi}}\ and\ \bibinfo {author} {\bibfnamefont {A.}~\bibnamefont
  {Kitaev}},\ }\bibfield  {title} {\bibinfo {title} {Fermionic quantum
  computation},\ }\href@noop {} {\bibfield  {journal} {\bibinfo  {journal}
  {Ann. of Phys.}\ }\textbf {\bibinfo {volume} {298}},\ \bibinfo {pages} {210}
  (\bibinfo {year} {2002})}\BibitemShut {NoStop}%
\bibitem [{\citenamefont {Bravyi}\ \emph {et~al.}(2017)\citenamefont {Bravyi},
  \citenamefont {Gambetta}, \citenamefont {Mezzacapo},\ and\ \citenamefont
  {Temme}}]{bravyi_2017}%
  \BibitemOpen
  \bibfield  {author} {\bibinfo {author} {\bibfnamefont {S.}~\bibnamefont
  {Bravyi}}, \bibinfo {author} {\bibfnamefont {J.~M.}\ \bibnamefont
  {Gambetta}}, \bibinfo {author} {\bibfnamefont {A.}~\bibnamefont
  {Mezzacapo}},\ and\ \bibinfo {author} {\bibfnamefont {K.}~\bibnamefont
  {Temme}},\ }\href {https://arxiv.org/abs/1701.08213} {\bibinfo {title}
  {Tapering off qubits to simulate fermionic hamiltonians.}} (\bibinfo {year}
  {2017})\BibitemShut {NoStop}%
\bibitem [{\citenamefont {Sim}\ \emph {et~al.}(2019)\citenamefont {Sim},
  \citenamefont {Johnson},\ and\ \citenamefont {Aspuru-Guzik}}]{guzik_2019}%
  \BibitemOpen
  \bibfield  {author} {\bibinfo {author} {\bibfnamefont {S.}~\bibnamefont
  {Sim}}, \bibinfo {author} {\bibfnamefont {P.~D.}\ \bibnamefont {Johnson}},\
  and\ \bibinfo {author} {\bibfnamefont {A.}~\bibnamefont {Aspuru-Guzik}},\
  }\bibfield  {title} {\bibinfo {title} {Expressibility and entangling
  capability of parameterized quantum circuits for hybrid quantum-classical
  algorithms},\ }\href {https://doi.org/https://doi.org/10.1002/qute.201900070}
  {\bibfield  {journal} {\bibinfo  {journal} {Advanced Quantum Technologies}\
  }\textbf {\bibinfo {volume} {2}},\ \bibinfo {pages} {1900070} (\bibinfo
  {year} {2019})}\BibitemShut {NoStop}%
\bibitem [{\citenamefont {Sokolov}\ \emph {et~al.}(2020)\citenamefont
  {Sokolov}, \citenamefont {Barkoutsos}, \citenamefont {Ollitrault},
  \citenamefont {Greenberg}, \citenamefont {Rice}, \citenamefont {Pistoia},\
  and\ \citenamefont {Tavernelli}}]{Sokolov2020}%
  \BibitemOpen
  \bibfield  {author} {\bibinfo {author} {\bibfnamefont {I.~O.}\ \bibnamefont
  {Sokolov}}, \bibinfo {author} {\bibfnamefont {P.~K.}\ \bibnamefont
  {Barkoutsos}}, \bibinfo {author} {\bibfnamefont {P.~J.}\ \bibnamefont
  {Ollitrault}}, \bibinfo {author} {\bibfnamefont {D.}~\bibnamefont
  {Greenberg}}, \bibinfo {author} {\bibfnamefont {J.}~\bibnamefont {Rice}},
  \bibinfo {author} {\bibfnamefont {M.}~\bibnamefont {Pistoia}},\ and\ \bibinfo
  {author} {\bibfnamefont {I.}~\bibnamefont {Tavernelli}},\ }\bibfield  {title}
  {\bibinfo {title} {Quantum orbital-optimized unitary coupled cluster methods
  in the strongly correlated regime: Can quantum algorithms outperform their
  classical equivalents?},\ }\href@noop {} {\bibfield  {journal} {\bibinfo
  {journal} {The Journal of chemical physics}\ }\textbf {\bibinfo {volume}
  {152}},\ \bibinfo {pages} {124107} (\bibinfo {year} {2020})}\BibitemShut
  {NoStop}%
\bibitem [{\citenamefont {Grimsley}\ \emph {et~al.}(2018)\citenamefont
  {Grimsley}, \citenamefont {Economou}, \citenamefont {Barnes},\ and\
  \citenamefont {Mayhall}}]{grimsley2018adapt}%
  \BibitemOpen
  \bibfield  {author} {\bibinfo {author} {\bibfnamefont {H.~R.}\ \bibnamefont
  {Grimsley}}, \bibinfo {author} {\bibfnamefont {S.~E.}\ \bibnamefont
  {Economou}}, \bibinfo {author} {\bibfnamefont {E.}~\bibnamefont {Barnes}},\
  and\ \bibinfo {author} {\bibfnamefont {N.~J.}\ \bibnamefont {Mayhall}},\
  }\bibfield  {title} {\bibinfo {title} {Adapt-vqe: An exact variational
  algorithm for fermionic simulations on a quantum computer},\ }\href@noop {}
  {\bibfield  {journal} {\bibinfo  {journal} {arXiv preprint arXiv:1812.11173}\
  } (\bibinfo {year} {2018})}\BibitemShut {NoStop}%
\bibitem [{\citenamefont {Tang}\ \emph {et~al.}(2021)\citenamefont {Tang},
  \citenamefont {Shkolnikov}, \citenamefont {Barron}, \citenamefont {Grimsley},
  \citenamefont {Mayhall}, \citenamefont {Barnes},\ and\ \citenamefont
  {Economou}}]{tang2021qubit}%
  \BibitemOpen
  \bibfield  {author} {\bibinfo {author} {\bibfnamefont {H.~L.}\ \bibnamefont
  {Tang}}, \bibinfo {author} {\bibfnamefont {V.}~\bibnamefont {Shkolnikov}},
  \bibinfo {author} {\bibfnamefont {G.~S.}\ \bibnamefont {Barron}}, \bibinfo
  {author} {\bibfnamefont {H.~R.}\ \bibnamefont {Grimsley}}, \bibinfo {author}
  {\bibfnamefont {N.~J.}\ \bibnamefont {Mayhall}}, \bibinfo {author}
  {\bibfnamefont {E.}~\bibnamefont {Barnes}},\ and\ \bibinfo {author}
  {\bibfnamefont {S.~E.}\ \bibnamefont {Economou}},\ }\bibfield  {title}
  {\bibinfo {title} {qubit-adapt-vqe: An adaptive algorithm for constructing
  hardware-efficient ans{\"a}tze on a quantum processor},\ }\href@noop {}
  {\bibfield  {journal} {\bibinfo  {journal} {PRX Quantum}\ }\textbf {\bibinfo
  {volume} {2}},\ \bibinfo {pages} {020310} (\bibinfo {year}
  {2021})}\BibitemShut {NoStop}%
\bibitem [{\citenamefont {Shen}\ \emph {et~al.}(2017)\citenamefont {Shen},
  \citenamefont {Zhang}, \citenamefont {Zhang}, \citenamefont {Zhang},
  \citenamefont {Yung},\ and\ \citenamefont {Kim}}]{shen_2017}%
  \BibitemOpen
  \bibfield  {author} {\bibinfo {author} {\bibfnamefont {Y.}~\bibnamefont
  {Shen}}, \bibinfo {author} {\bibfnamefont {X.}~\bibnamefont {Zhang}},
  \bibinfo {author} {\bibfnamefont {S.}~\bibnamefont {Zhang}}, \bibinfo
  {author} {\bibfnamefont {J.-N.}\ \bibnamefont {Zhang}}, \bibinfo {author}
  {\bibfnamefont {M.-H.}\ \bibnamefont {Yung}},\ and\ \bibinfo {author}
  {\bibfnamefont {K.}~\bibnamefont {Kim}},\ }\bibfield  {title} {\bibinfo
  {title} {Quantum implementation of the unitary coupled cluster for simulating
  molecular electronic structure},\ }\href@noop {} {\bibfield  {journal}
  {\bibinfo  {journal} {Physical Review A}\ }\textbf {\bibinfo {volume} {95}},\
  \bibinfo {pages} {020501} (\bibinfo {year} {2017})}\BibitemShut {NoStop}%
\bibitem [{\citenamefont {Morse}(1929)}]{morse_1929}%
  \BibitemOpen
  \bibfield  {author} {\bibinfo {author} {\bibfnamefont {P.~M.}\ \bibnamefont
  {Morse}},\ }\bibfield  {title} {\bibinfo {title} {Diatomic molecules
  according to the wave mechanics. ii. vibrational levels},\ }\href@noop {}
  {\bibfield  {journal} {\bibinfo  {journal} {Physical Review}\ }\textbf
  {\bibinfo {volume} {34}},\ \bibinfo {pages} {57} (\bibinfo {year}
  {1929})}\BibitemShut {NoStop}%
\bibitem [{\citenamefont {Abraham}\ \emph {et~al.}(2019)\citenamefont {Abraham}
  \emph {et~al.}}]{Qiskit}%
  \BibitemOpen
  \bibfield  {author} {\bibinfo {author} {\bibfnamefont {H.}~\bibnamefont
  {Abraham}} \emph {et~al.},\ }\href {https://doi.org/10.5281/zenodo.2562110}
  {\bibinfo {title} {Qiskit: An open-source framework for quantum computing}},\
  \bibinfo {howpublished} {DOI: 10.5281/zenodo.2562110} (\bibinfo {year}
  {2019})\BibitemShut {NoStop}%
\bibitem [{\citenamefont {Lavrijsen}\ \emph {et~al.}(2020)\citenamefont
  {Lavrijsen}, \citenamefont {Tudor}, \citenamefont {M{\"u}ller}, \citenamefont
  {Iancu},\ and\ \citenamefont {de~Jong}}]{lavrijsen2020classical}%
  \BibitemOpen
  \bibfield  {author} {\bibinfo {author} {\bibfnamefont {W.}~\bibnamefont
  {Lavrijsen}}, \bibinfo {author} {\bibfnamefont {A.}~\bibnamefont {Tudor}},
  \bibinfo {author} {\bibfnamefont {J.}~\bibnamefont {M{\"u}ller}}, \bibinfo
  {author} {\bibfnamefont {C.}~\bibnamefont {Iancu}},\ and\ \bibinfo {author}
  {\bibfnamefont {W.}~\bibnamefont {de~Jong}},\ }\bibfield  {title} {\bibinfo
  {title} {Classical optimizers for noisy intermediate-scale quantum devices},\
  }in\ \href@noop {} {\emph {\bibinfo {booktitle} {2020 IEEE International
  Conference on Quantum Computing and Engineering (QCE)}}}\ (\bibinfo
  {organization} {IEEE},\ \bibinfo {year} {2020})\ pp.\ \bibinfo {pages}
  {267--277}\BibitemShut {NoStop}%
\bibitem [{\citenamefont {Harwood}\ \emph {et~al.}()\citenamefont {Harwood},
  \citenamefont {Trenev}, \citenamefont {Stober}, \citenamefont {Barkoutsos},
  \citenamefont {Gujarati},\ and\ \citenamefont {Mostame}}]{harwood_2021}%
  \BibitemOpen
  \bibfield  {author} {\bibinfo {author} {\bibfnamefont {S.~M.}\ \bibnamefont
  {Harwood}}, \bibinfo {author} {\bibfnamefont {D.}~\bibnamefont {Trenev}},
  \bibinfo {author} {\bibfnamefont {S.~T.}\ \bibnamefont {Stober}}, \bibinfo
  {author} {\bibfnamefont {P.}~\bibnamefont {Barkoutsos}}, \bibinfo {author}
  {\bibfnamefont {T.~P.}\ \bibnamefont {Gujarati}},\ and\ \bibinfo {author}
  {\bibfnamefont {S.}~\bibnamefont {Mostame}},\ }\href
  {https://arxiv.org/abs/2102.02875} {\bibinfo {title} {Improving the
  variational quantum eigensolver using variational adiabatic quantum
  computing}}\BibitemShut {NoStop}%
\bibitem [{\citenamefont {McClean}\ \emph {et~al.}(2016)\citenamefont
  {McClean}, \citenamefont {Romero}, \citenamefont {Babbush},\ and\
  \citenamefont {Aspuru-Guzik}}]{mcclean2016theory}%
  \BibitemOpen
  \bibfield  {author} {\bibinfo {author} {\bibfnamefont {J.~R.}\ \bibnamefont
  {McClean}}, \bibinfo {author} {\bibfnamefont {J.}~\bibnamefont {Romero}},
  \bibinfo {author} {\bibfnamefont {R.}~\bibnamefont {Babbush}},\ and\ \bibinfo
  {author} {\bibfnamefont {A.}~\bibnamefont {Aspuru-Guzik}},\ }\bibfield
  {title} {\bibinfo {title} {The theory of variational hybrid quantum-classical
  algorithms},\ }\href@noop {} {\bibfield  {journal} {\bibinfo  {journal} {New
  Journal of Physics}\ }\textbf {\bibinfo {volume} {18}},\ \bibinfo {pages}
  {023023} (\bibinfo {year} {2016})}\BibitemShut {NoStop}%
\bibitem [{\citenamefont {Wilson}\ \emph {et~al.}(1955)\citenamefont {Wilson},
  \citenamefont {Decius},\ and\ \citenamefont {Cross}}]{book_wilson}%
  \BibitemOpen
  \bibfield  {author} {\bibinfo {author} {\bibfnamefont {B.~E.}\ \bibnamefont
  {Wilson}}, \bibinfo {author} {\bibfnamefont {J.~C.}\ \bibnamefont {Decius}},\
  and\ \bibinfo {author} {\bibfnamefont {P.~C.}\ \bibnamefont {Cross}},\
  }\href@noop {} {\emph {\bibinfo {title} {Molecular Vibrations: The Theory of
  Infrared and Raman Vibrational Spectra}}}\ (\bibinfo  {publisher}
  {McGraw-Hill Book Complany Inc.},\ \bibinfo {address} {New York, NY},\
  \bibinfo {year} {1955})\BibitemShut {NoStop}%
\bibitem [{\citenamefont {McQuarrie}(2000)}]{book_mcquarrie}%
  \BibitemOpen
  \bibfield  {author} {\bibinfo {author} {\bibfnamefont {D.~A.}\ \bibnamefont
  {McQuarrie}},\ }\href@noop {} {\emph {\bibinfo {title} {Statistical
  Mechanics}}}\ (\bibinfo  {publisher} {University Science Books},\ \bibinfo
  {address} {Sausalito, CA},\ \bibinfo {year} {2000})\BibitemShut {NoStop}%
\bibitem [{\citenamefont {Stober}\ \emph {et~al.}(2020)\citenamefont {Stober},
  \citenamefont {Harwood}, \citenamefont {Greenberg}, \citenamefont {Gujarati},
  \citenamefont {Mostame}, \citenamefont {Raman},\ and\ \citenamefont
  {Trenev}}]{stober_2020}%
  \BibitemOpen
  \bibfield  {author} {\bibinfo {author} {\bibfnamefont {S.~T.}\ \bibnamefont
  {Stober}}, \bibinfo {author} {\bibfnamefont {S.~M.}\ \bibnamefont {Harwood}},
  \bibinfo {author} {\bibfnamefont {D.}~\bibnamefont {Greenberg}}, \bibinfo
  {author} {\bibfnamefont {T.~P.}\ \bibnamefont {Gujarati}}, \bibinfo {author}
  {\bibfnamefont {S.}~\bibnamefont {Mostame}}, \bibinfo {author} {\bibfnamefont
  {S.}~\bibnamefont {Raman}},\ and\ \bibinfo {author} {\bibfnamefont
  {D.}~\bibnamefont {Trenev}},\ }\href {https://arxiv.org/abs/2003.02303}
  {\bibinfo {title} {Computing thermodynamic observables on noisy quantum
  computers with chemical accuracy}} (\bibinfo {year} {2020})\BibitemShut
  {NoStop}%
\bibitem [{\citenamefont {Dunning}(1989)}]{dunning_1989}%
  \BibitemOpen
  \bibfield  {author} {\bibinfo {author} {\bibfnamefont {T.~H.}\ \bibnamefont
  {Dunning}},\ }\bibfield  {title} {\bibinfo {title} {Gaussian basis sets for
  use in correlated molecular calculations. i. the atoms boron through neon and
  hydrogen},\ }\href@noop {} {\bibfield  {journal} {\bibinfo  {journal} {The
  Journal of Chemical Physics}\ }\textbf {\bibinfo {volume} {90}},\ \bibinfo
  {pages} {1007} (\bibinfo {year} {1989})}\BibitemShut {NoStop}%
\bibitem [{\citenamefont {Parrish}\ \emph {et~al.}(2017)\citenamefont
  {Parrish}, \citenamefont {Burns}, \citenamefont {Smith}, \citenamefont
  {Simmonett}, \citenamefont {DePrince}, \citenamefont {Hohenstein},
  \citenamefont {Bozkaya}, \citenamefont {Sokolov}, \citenamefont {Di~Remigio},
  \citenamefont {Richard}, \citenamefont {Gonthier}, \citenamefont {James},
  \citenamefont {McAlexander}, \citenamefont {Kumar}, \citenamefont {Saitow},
  \citenamefont {Wang}, \citenamefont {Pritchard}, \citenamefont {Verma},
  \citenamefont {Schaefer}, \citenamefont {Patkowski}, \citenamefont {King},
  \citenamefont {Valeev}, \citenamefont {Evangelista}, \citenamefont {Turney},
  \citenamefont {Crawford},\ and\ \citenamefont {Sherrill}}]{psi4}%
  \BibitemOpen
  \bibfield  {author} {\bibinfo {author} {\bibfnamefont {R.~M.}\ \bibnamefont
  {Parrish}}, \bibinfo {author} {\bibfnamefont {L.~A.}\ \bibnamefont {Burns}},
  \bibinfo {author} {\bibfnamefont {D.~G.~A.}\ \bibnamefont {Smith}}, \bibinfo
  {author} {\bibfnamefont {A.~C.}\ \bibnamefont {Simmonett}}, \bibinfo {author}
  {\bibfnamefont {A.~E.}\ \bibnamefont {DePrince}}, \bibinfo {author}
  {\bibfnamefont {E.~G.}\ \bibnamefont {Hohenstein}}, \bibinfo {author}
  {\bibfnamefont {U.}~\bibnamefont {Bozkaya}}, \bibinfo {author} {\bibfnamefont
  {A.~Y.}\ \bibnamefont {Sokolov}}, \bibinfo {author} {\bibfnamefont
  {R.}~\bibnamefont {Di~Remigio}}, \bibinfo {author} {\bibfnamefont {R.~M.}\
  \bibnamefont {Richard}}, \bibinfo {author} {\bibfnamefont {J.~F.}\
  \bibnamefont {Gonthier}}, \bibinfo {author} {\bibfnamefont {A.~M.}\
  \bibnamefont {James}}, \bibinfo {author} {\bibfnamefont {H.~R.}\ \bibnamefont
  {McAlexander}}, \bibinfo {author} {\bibfnamefont {A.}~\bibnamefont {Kumar}},
  \bibinfo {author} {\bibfnamefont {M.}~\bibnamefont {Saitow}}, \bibinfo
  {author} {\bibfnamefont {X.}~\bibnamefont {Wang}}, \bibinfo {author}
  {\bibfnamefont {B.~P.}\ \bibnamefont {Pritchard}}, \bibinfo {author}
  {\bibfnamefont {P.}~\bibnamefont {Verma}}, \bibinfo {author} {\bibfnamefont
  {H.~F.}\ \bibnamefont {Schaefer}}, \bibinfo {author} {\bibfnamefont
  {K.}~\bibnamefont {Patkowski}}, \bibinfo {author} {\bibfnamefont {R.~A.}\
  \bibnamefont {King}}, \bibinfo {author} {\bibfnamefont {E.~F.}\ \bibnamefont
  {Valeev}}, \bibinfo {author} {\bibfnamefont {F.~A.}\ \bibnamefont
  {Evangelista}}, \bibinfo {author} {\bibfnamefont {J.~M.}\ \bibnamefont
  {Turney}}, \bibinfo {author} {\bibfnamefont {T.~D.}\ \bibnamefont
  {Crawford}},\ and\ \bibinfo {author} {\bibfnamefont {C.~D.}\ \bibnamefont
  {Sherrill}},\ }\bibfield  {title} {\bibinfo {title} {Psi4 1.1: An open-source
  electronic structure program emphasizing automation, advanced libraries, and
  interoperability},\ }\href@noop {} {\bibfield  {journal} {\bibinfo  {journal}
  {Journal of Chemical Theory and Computation}\ }\textbf {\bibinfo {volume}
  {13}},\ \bibinfo {pages} {3185} (\bibinfo {year} {2017})}\BibitemShut
  {NoStop}%
\bibitem [{ccc()}]{cccbdb}%
  \BibitemOpen
  \href {http://cccbdb.nist.gov/} {\bibinfo {title} {Nist computational
  chemistry comparison and benchmark database}},\ \bibinfo {howpublished}
  {Release 21, August 2020, Editor: Russell D. Johnson III},\ \bibinfo {note}
  {nIST Standard Reference Database Number 101}\BibitemShut {NoStop}%
\bibitem [{\citenamefont {Chase}(1998)}]{nist_data}%
  \BibitemOpen
  \bibfield  {author} {\bibinfo {author} {\bibfnamefont {M.}~\bibnamefont
  {Chase}, \bibfnamefont {Jr.}},\ }\href@noop {} {\emph {\bibinfo {title}
  {NIST-JANAF Themochemical Tables, Fourth Edition}}}\ (\bibinfo  {publisher}
  {J. Phys. Chem. Ref. Data, Monograph 9},\ \bibinfo {year} {1998})\BibitemShut
  {NoStop}%
\bibitem [{\citenamefont {Popovas}\ and\ \citenamefont
  {Jorgensen}(2016)}]{popovas_2016}%
  \BibitemOpen
  \bibfield  {author} {\bibinfo {author} {\bibfnamefont {A.}~\bibnamefont
  {Popovas}}\ and\ \bibinfo {author} {\bibfnamefont {U.~G.}\ \bibnamefont
  {Jorgensen}},\ }\bibfield  {title} {\bibinfo {title} {I. improved partition
  functions and thermodynamic quantities for normal, equilibrium, and ortho and
  para molecular hydrogen},\ }\href@noop {} {\bibfield  {journal} {\bibinfo
  {journal} {Astronomy and Astrophysics}\ }\textbf {\bibinfo {volume} {595}},\
  \bibinfo {pages} {A130} (\bibinfo {year} {2016})}\BibitemShut {NoStop}%
\bibitem [{\citenamefont {Schuld}\ \emph {et~al.}(2019)\citenamefont {Schuld},
  \citenamefont {Bergholm}, \citenamefont {Gogolin}, \citenamefont {Izaac},\
  and\ \citenamefont {Killoran}}]{schuld2019evaluating}%
  \BibitemOpen
  \bibfield  {author} {\bibinfo {author} {\bibfnamefont {M.}~\bibnamefont
  {Schuld}}, \bibinfo {author} {\bibfnamefont {V.}~\bibnamefont {Bergholm}},
  \bibinfo {author} {\bibfnamefont {C.}~\bibnamefont {Gogolin}}, \bibinfo
  {author} {\bibfnamefont {J.}~\bibnamefont {Izaac}},\ and\ \bibinfo {author}
  {\bibfnamefont {N.}~\bibnamefont {Killoran}},\ }\bibfield  {title} {\bibinfo
  {title} {Evaluating analytic gradients on quantum hardware},\ }\href@noop {}
  {\bibfield  {journal} {\bibinfo  {journal} {Physical Review A}\ }\textbf
  {\bibinfo {volume} {99}},\ \bibinfo {pages} {032331} (\bibinfo {year}
  {2019})}\BibitemShut {NoStop}%
\bibitem [{\citenamefont {Crooks}(2019)}]{crooks_2019}%
  \BibitemOpen
  \bibfield  {author} {\bibinfo {author} {\bibfnamefont {G.~E.}\ \bibnamefont
  {Crooks}},\ }\bibfield  {title} {\bibinfo {title} {Gradients of parameterized
  quantum gates using the parameter-shift rule and gate decomposition},\
  }\href@noop {} {\bibfield  {journal} {\bibinfo  {journal} {arXiv preprint
  arXiv:1905.13311}\ } (\bibinfo {year} {2019})}\BibitemShut {NoStop}%
\bibitem [{\citenamefont {Mitarai}\ \emph {et~al.}(2020)\citenamefont
  {Mitarai}, \citenamefont {Nakagawa},\ and\ \citenamefont
  {Mizukami}}]{mitarai2020theory}%
  \BibitemOpen
  \bibfield  {author} {\bibinfo {author} {\bibfnamefont {K.}~\bibnamefont
  {Mitarai}}, \bibinfo {author} {\bibfnamefont {Y.~O.}\ \bibnamefont
  {Nakagawa}},\ and\ \bibinfo {author} {\bibfnamefont {W.}~\bibnamefont
  {Mizukami}},\ }\bibfield  {title} {\bibinfo {title} {Theory of analytical
  energy derivatives for the variational quantum eigensolver},\ }\href@noop {}
  {\bibfield  {journal} {\bibinfo  {journal} {Physical Review Research}\
  }\textbf {\bibinfo {volume} {2}},\ \bibinfo {pages} {013129} (\bibinfo {year}
  {2020})}\BibitemShut {NoStop}%
\bibitem [{\citenamefont {Bertsekas}\ and\ \citenamefont
  {Tsitsiklis}(2000)}]{bertsekas_2000}%
  \BibitemOpen
  \bibfield  {author} {\bibinfo {author} {\bibfnamefont {D.~P.}\ \bibnamefont
  {Bertsekas}}\ and\ \bibinfo {author} {\bibfnamefont {J.~N.}\ \bibnamefont
  {Tsitsiklis}},\ }\bibfield  {title} {\bibinfo {title} {Gradient convergence
  in gradient methods with errors},\ }\href@noop {} {\bibfield  {journal}
  {\bibinfo  {journal} {SIAM Journal on Optimization}\ }\textbf {\bibinfo
  {volume} {10}},\ \bibinfo {pages} {627} (\bibinfo {year} {2000})}\BibitemShut
  {NoStop}%
\bibitem [{\citenamefont {Larsson}\ and\ \citenamefont
  {Thomee}(2008)}]{larsson2008partial}%
  \BibitemOpen
  \bibfield  {author} {\bibinfo {author} {\bibfnamefont {S.}~\bibnamefont
  {Larsson}}\ and\ \bibinfo {author} {\bibfnamefont {V.}~\bibnamefont
  {Thomee}},\ }\href@noop {} {\emph {\bibinfo {title} {Partial Differential
  Equations with Numerical Methods}}},\ Texts in Applied Mathematics\ (\bibinfo
   {publisher} {Springer Berlin Heidelberg},\ \bibinfo {year}
  {2008})\BibitemShut {NoStop}%
\bibitem [{\citenamefont {Virtanen}\ \emph {et~al.}(2020)\citenamefont
  {Virtanen}, \citenamefont {Gommers}, \citenamefont {Oliphant}, \citenamefont
  {Haberland}, \citenamefont {Reddy}, \citenamefont {Cournapeau}, \citenamefont
  {Burovski}, \citenamefont {Peterson}, \citenamefont {Weckesser},
  \citenamefont {Bright}, \citenamefont {{van der Walt}}, \citenamefont
  {Brett}, \citenamefont {Wilson}, \citenamefont {Millman}, \citenamefont
  {Mayorov}, \citenamefont {Nelson}, \citenamefont {Jones}, \citenamefont
  {Kern}, \citenamefont {Larson}, \citenamefont {Carey}, \citenamefont {Polat},
  \citenamefont {Feng}, \citenamefont {Moore}, \citenamefont {{VanderPlas}},
  \citenamefont {Laxalde}, \citenamefont {Perktold}, \citenamefont {Cimrman},
  \citenamefont {Henriksen}, \citenamefont {Quintero}, \citenamefont {Harris},
  \citenamefont {Archibald}, \citenamefont {Ribeiro}, \citenamefont
  {Pedregosa}, \citenamefont {{van Mulbregt}},\ and\ \citenamefont {{SciPy 1.0
  Contributors}}}]{2020SciPy-NMeth}%
  \BibitemOpen
  \bibfield  {author} {\bibinfo {author} {\bibfnamefont {P.}~\bibnamefont
  {Virtanen}}, \bibinfo {author} {\bibfnamefont {R.}~\bibnamefont {Gommers}},
  \bibinfo {author} {\bibfnamefont {T.~E.}\ \bibnamefont {Oliphant}}, \bibinfo
  {author} {\bibfnamefont {M.}~\bibnamefont {Haberland}}, \bibinfo {author}
  {\bibfnamefont {T.}~\bibnamefont {Reddy}}, \bibinfo {author} {\bibfnamefont
  {D.}~\bibnamefont {Cournapeau}}, \bibinfo {author} {\bibfnamefont
  {E.}~\bibnamefont {Burovski}}, \bibinfo {author} {\bibfnamefont
  {P.}~\bibnamefont {Peterson}}, \bibinfo {author} {\bibfnamefont
  {W.}~\bibnamefont {Weckesser}}, \bibinfo {author} {\bibfnamefont
  {J.}~\bibnamefont {Bright}}, \bibinfo {author} {\bibfnamefont {S.~J.}\
  \bibnamefont {{van der Walt}}}, \bibinfo {author} {\bibfnamefont
  {M.}~\bibnamefont {Brett}}, \bibinfo {author} {\bibfnamefont
  {J.}~\bibnamefont {Wilson}}, \bibinfo {author} {\bibfnamefont {K.~J.}\
  \bibnamefont {Millman}}, \bibinfo {author} {\bibfnamefont {N.}~\bibnamefont
  {Mayorov}}, \bibinfo {author} {\bibfnamefont {A.~R.~J.}\ \bibnamefont
  {Nelson}}, \bibinfo {author} {\bibfnamefont {E.}~\bibnamefont {Jones}},
  \bibinfo {author} {\bibfnamefont {R.}~\bibnamefont {Kern}}, \bibinfo {author}
  {\bibfnamefont {E.}~\bibnamefont {Larson}}, \bibinfo {author} {\bibfnamefont
  {C.~J.}\ \bibnamefont {Carey}}, \bibinfo {author} {\bibfnamefont
  {{\.I}.}~\bibnamefont {Polat}}, \bibinfo {author} {\bibfnamefont
  {Y.}~\bibnamefont {Feng}}, \bibinfo {author} {\bibfnamefont {E.~W.}\
  \bibnamefont {Moore}}, \bibinfo {author} {\bibfnamefont {J.}~\bibnamefont
  {{VanderPlas}}}, \bibinfo {author} {\bibfnamefont {D.}~\bibnamefont
  {Laxalde}}, \bibinfo {author} {\bibfnamefont {J.}~\bibnamefont {Perktold}},
  \bibinfo {author} {\bibfnamefont {R.}~\bibnamefont {Cimrman}}, \bibinfo
  {author} {\bibfnamefont {I.}~\bibnamefont {Henriksen}}, \bibinfo {author}
  {\bibfnamefont {E.~A.}\ \bibnamefont {Quintero}}, \bibinfo {author}
  {\bibfnamefont {C.~R.}\ \bibnamefont {Harris}}, \bibinfo {author}
  {\bibfnamefont {A.~M.}\ \bibnamefont {Archibald}}, \bibinfo {author}
  {\bibfnamefont {A.~H.}\ \bibnamefont {Ribeiro}}, \bibinfo {author}
  {\bibfnamefont {F.}~\bibnamefont {Pedregosa}}, \bibinfo {author}
  {\bibfnamefont {P.}~\bibnamefont {{van Mulbregt}}},\ and\ \bibinfo {author}
  {\bibnamefont {{SciPy 1.0 Contributors}}},\ }\bibfield  {title} {\bibinfo
  {title} {{{SciPy} 1.0: Fundamental Algorithms for Scientific Computing in
  Python}},\ }\href {https://doi.org/10.1038/s41592-019-0686-2} {\bibfield
  {journal} {\bibinfo  {journal} {Nature Methods}\ }\textbf {\bibinfo {volume}
  {17}},\ \bibinfo {pages} {261} (\bibinfo {year} {2020})}\BibitemShut
  {NoStop}%
\end{thebibliography}

%

\end{document}